\documentclass[11pt]{article}
\usepackage{graphicx}
\usepackage{float}
\usepackage[utf8]{inputenc}
\usepackage[font=small]{caption}
\usepackage{xcolor}
\usepackage{soul} 

\usepackage{indentfirst} 
\usepackage{amsmath}
\usepackage{amssymb}
\usepackage{bm}
\usepackage{mathtools}
\usepackage{array}
\usepackage{blkarray}
\usepackage{booktabs}
\usepackage{dsfont}

\usepackage{geometry}
\usepackage{hyperref}
\hypersetup{
    colorlinks,
    linkcolor={red!50!black},
    citecolor={blue!50!black},
    urlcolor={blue!80!black}
}

\usepackage[section]{placeins} 
\usepackage[backend=biber,
            style=apa,
            maxcitenames=2,
            mincitenames=1,
            apamaxprtauth=6,
            uniquename=false,
            natbib=true,
            date=year,
            sortcites=ynt,
            url=false,
            doi=false]{biblatex} 
\usepackage[title]{appendix}

\usepackage{tabularx,ragged2e,booktabs}
\usepackage{authblk}
\usepackage{tcolorbox} 
\usepackage[nodisplayskipstretch]{setspace}
\usepackage[ruled,linesnumbered,noend]{algorithm2e}

\AtBeginEnvironment{algorithm}{\setstretch{1}}

\begin{document}

\title{Haplotype frequency inference from pooled genetic data with a latent multinomial model}

\author[1]{Yong See Foo}
\author[1]{Jennifer A. Flegg}
\affil[1]{School of Mathematics and Statistics, The University of Melbourne, Parkville, Australia}
\date{}
\setcounter{Maxaffil}{0}
\renewcommand\Affilfont{\itshape\small}
\maketitle
\begin{abstract}
 In genetic studies, haplotype data provide more refined information than data about separate genetic markers. However, large-scale studies that genotype hundreds to thousands of individuals may only provide results of pooled data, where only the total allele counts of each marker in each pool are reported. Methods for inferring haplotype frequencies from pooled genetic data that scale well with pool size rely on a normal approximation, which we observe to produce unreliable inference when applied to real data. We illustrate cases where the approximation breaks down, due to the normal covariance matrix being near-singular. As an alternative to approximate methods, in this paper we propose exact methods to infer haplotype frequencies from pooled genetic data based on a latent multinomial model, where the observed allele counts are considered integer combinations of latent, unobserved haplotype counts. One of our methods, latent count sampling via Markov bases, achieves approximately linear runtime with respect to pool size. Our exact methods produce more accurate inference over existing approximate methods for synthetic data and for data based on haplotype information from the 1000 Genomes Project. We also demonstrate how our methods can be applied to time-series of pooled genetic data, as a proof of concept of how our methods are relevant to more complex hierarchical settings, such as spatiotemporal models. 
\end{abstract}


\section{Introduction}\label{sec:intro}

In large-scale genetic studies, individuals are genotyped at multiple genetic markers, often for the purpose of studying genetic association. These markers may exhibit mutational change, the most common being single nucleotide polymorphisms (SNPs), where nucleotide variations of single bases are called alleles~\parencite{wright_genetic_2005}. In order to reduce genotyping costs, DNA data of up to hundreds of individuals may be pooled into several groups before genotyping, instead of determining the sequence of alleles for each individual separately. As a result, we only retain the allele counts of each SNP for each pool, and lose information about the configuration of alleles over SNPs. Apart from data that is pooled during genotyping, pooled results can also come from studies where data is partially reported. Even if individual-level genotyping is performed, the results may be summarised such that only pooled data over individual markers is available.

SNPs that are close to each other are often correlated, resulting in limited variation of haplotypes (combinations of SNP alleles in a genetic region)~\parencite{wright_genetic_2005}. Rather than analysing SNPs separately, haplotypes provide finer information when associating genetic data to phenotypes (observable traits of an organism)~\parencite{tam_benefits_2019}. In this paper, we address the statistical inverse problem of inferring the frequencies of haplotypes given pooled genetic data, i.e. pooled allele counts of each marker. Some previous methods rely on enumerating all possible haplotype assignments~\parencite{ito_estimation_2003,kirkpatrick_haplopool_2007,iliadis_fast_2012}, but they are only applicable to small pool sizes ($\le 20$ haplotype samples per pool). As genetic studies can have up to hundreds of samples per pool~\citep{zhang_poool_2008}, methods that scale well with pool size are needed. An example of such an approach is sparse optimisation, which solves to find haplotype frequency vectors that are compatible with the observed allele frequencies, and have only a few nonzero entries~\citep{jajamovich_maximum-parsimony_2013,zhou_cshap_2019}. This reflects the reality that given a sequence of markers, only a few out of the exponentially many possible haplotypes are present in a population~\parencite{patil_blocks_2001}. However, it is not straightforward to quantify uncertainties of the inferred frequencies, which impedes downstream statistical inference. There are also statistical methods that avoid enumerating haplotype assignments by using a normal approximation~\citep{zhang_poool_2008,kuk_computationally_2009,pirinen_estimating_2009}, thereby achieving computational runtimes that are fairly insensitive to pool size. The authors claim that the error introduced by the normal approximation is negligible for large pool sizes due to the central limit theorem. In particular, it is the \emph{multivariate} central limit theorem that applies, which requires the covariance matrix to be non-singular for the probability density to be finite. However, some haplotype frequencies can give rise to singular covariance matrices, which causes the normal approximation to break down. This issue has not been previously acknowledged in the works that use the normal approximation, bringing the reliability of their methods into question.

To address this issue, we develop two exact Bayesian methods to perform haplotype frequency estimation for large pools of genetic data, in order to test whether approximate methods give results that are comparable to exact methods. Our first method enumerates all haplotype assignments using a branch-and-bound algorithm, whereas the second method treats the counts of each haplotype for each pool as latent variables to be inferred. Although the first method does not scale well with pool size, we demonstrate its utility for an example over 8 haplotypes with pools up to 100 samples each. On the other hand, the runtime of our second method scales well; its runtime is approximately linear with respect to pool size. To scale our methods with the number of markers, we incorporate partition ligation~\parencite{niu_bayesian_2002}. When dealing with a long sequence of markers, the partition ligation procedure estimates frequencies of partial haplotypes over short segments of markers, and subsequently stitches the segments back in a recursive manner. This avoids having to perform inference on too many haplotypes simultaneously.

We formulate both of our methods under a \emph{latent multinomial} framework, where the counts of each haplotype for each pool are modelled as latent multinomial counts that are unobserved, and the haplotype frequencies are modelled as multinomial probabilities. The observed allele counts of each marker in each pool are subsequently modelled as integer combinations of the latent counts. For our first exact method, we marginalise out these latent counts exactly, resulting in an enumeration-based approach. In the analysis of mark-recapture data~\citep{link_uncovering_2010,schofield_connecting_2015}, where animals are captured and released multiple times, the latent multinomial model has been used to handle the fact that the capture history of each individual is only partially observable and potentially erroneous. The authors treat the latent counts as discrete parameters, and sample them with a Markov chain Monte Carlo (MCMC) scheme. This is an exact inference method for the latent multinomial model, which we adopt for our second exact method.

We compare the performance of our two exact methods with an approximate counterpart of our first method, along with approximate methods from literature; the different methods are detailed in Section~\ref{methods}. To carry out the comparisons, we apply these methods to a simulation study based on synthetic data, and an example based on data from the 1000 Genomes Project~\parencite{the_1000_genomes_project_consortium_global_2015} in Section~\ref{sec:results}. We demonstrate that our exact methods produce more reliable inference without resorting to approximation, at the cost of longer computational runtimes. We also illustrate how our proposed methods can be applied in hierarchical settings e.g. time-series modelling or spatiotemporal modelling, which has not been previously done for haplotype frequency estimation on pooled genetic data. Finally, we discuss the implications of our findings in Section~\ref{discussion}.
 
\section{Methods}\label{methods}

We aim to perform inference on population haplotype frequencies over $M$ biallelic markers (i.e. each marker can be one of two possible alleles). For each marker, we represent the allele that occurs with higher frequency (major allele) as 0, and the allele that occurs with lower frequency (minor allele) as 1. A haplotype is represented by a string of $M$ binary digits. Suppose we have a set of $H$ \emph{input haplotypes}, where ${2\le H \le 2^M}$, such that the haplotypes present in the population is a subset of the input haplotypes. The genetic data is divided into $N$ pools, where pool~$i$ consists of $n_i$ haplotype samples for ${i=1,\ldots,N}$. Let ${\mathbf{z}_i \coloneqq (z_{i1},\ldots,z_{iH})}$ denote the number of occurrences of each input haplotype in pool~$i$ for ${i=1,\ldots,N}$. Assuming that the haplotype samples are unrelated, we have that
\begin{equation}
    \mathbf{z}_i \vert \mathbf{p} \sim \mathrm{Mult}(n_i; \mathbf{p}),
    \label{eq:multinom}
\end{equation}
where $\mathbf{p} \coloneqq (p_1,\ldots,p_H)$ are the population haplotype frequencies. Since some input haplotypes may be absent from the population, we allow the entries of $\mathbf{p}$ to be zero.

However, we do not directly observe the haplotype counts $\mathbf{z}_i$. Instead, for each pool~$i$, we observe the numbers of samples belonging to various subsets of haplotypes, and treat $\mathbf{z}_i$ as latent counts. For example, suppose that there are ${M=3}$ markers and we have prior knowledge to exclude haplotype~$111$ from the input haplotypes. Observing the number of samples with a minor allele at the first marker is then equivalent to observing the number of samples whose haplotypes are in the subset $\{100, 101, 110\}$. Suppose for each pool~$i$, we observe $R_i$ counts arranged as a vector ${\mathbf{y}_i \coloneqq (y_{i1},\ldots,y_{iR_i})}$. The observed count vector $\mathbf{y}_i$ is related to the latent count vector $\mathbf{z}_i$ through a ${R_i\times H}$ binary matrix $\mathbf{A}_i$ by the linear system ${\mathbf{y}_i = \mathbf{A}_i\mathbf{z}_i}$. The matrices $\mathbf{A}_i$ are called \emph{configuration matrices}. Each row of a configuration matrix is determined by the haplotypes associated with the corresponding observed count. Continuing the previous example, if for each marker we observe the number of samples with a minor allele, then each column of the configuration matrix matches the binary representation of the corresponding haplotype. In general, the configuration matrix may be different for each pool, depending on the subsets of haplotypes accounted by the observed counts for that pool. This is relevant for meta-analyses, where the genetic markers that each study reports on are not all the same.

The distribution of $\mathbf{y}_i\vert \mathbf{p}$ is known as a \emph{latent multinomial distribution}~\parencite{link_uncovering_2010}. A direct calculation of the probability mass function $p(\mathbf{y}_i\vert \mathbf{p})$ is  requires finding all latent counts $\mathbf{z}_i$ that are compatible with the observed counts $\mathbf{y}_i$, i.e. solving the system
\begin{align}
    \mathbf{A}_i\mathbf{z}_i &= \mathbf{y}_i, \label{eq:linear_sys}\\
    z_{i1}+\cdots+z_{iH} &= n_i, \label{eq:sum_to_n}\\
    z_{ih} &\ge 0 \quad\text{for }h=1,\ldots,H \label{eq:nonneg}
\end{align}
over nonnegative integers $z_{i1},\ldots,z_{iH}$. Solving the system \eqref{eq:linear_sys}--\eqref{eq:nonneg} is considered computationally intensive for large pool sizes $n_i$~\parencite{zhang_poool_2008,kuk_computationally_2009}. We review two methods in the literature that avoid this computation by using a normal approximation, and propose alternative approaches of handling $p(\mathbf{y}_i\vert \mathbf{p})$ without resorting to approximations.

\subsection{Existing approaches}\label{sec:existing}

According to the central limit theorem, the observed counts $\mathbf{y}_i$ are asymptotically normally distributed as the pool size $n_i$ increases~\parencite{zhang_poool_2008}. In particular, the distribution ${\mathbf{y}_i\vert \mathbf{p}}$ is approximately multivariate normal:
\begin{equation}\label{eq:mn_approx}
    \mathbf{y}_i\vert \mathbf{p} \approx \mathcal{N}\!\left(n_i\mathbf{A}_i\mathbf{p}, {n_i\mathbf{A}_i(\mathrm{diag}(\mathbf{p})-\mathbf{p}\mathbf{p}^T)\mathbf{A}_i^T}\right),
\end{equation}
given that the covariance matrix ${n_i\mathbf{A}_i(\mathrm{diag}(\mathbf{p})-\mathbf{p}\mathbf{p}^T)\mathbf{A}_i^T}$ is non-singular. \textcite{kuk_computationally_2009} proposed an approximate expectation-maximisation (AEM) algorithm to approximate the maximum likelihood estimate of $\mathbf{p}$, where the likelihood is approximated based on \eqref{eq:mn_approx}. The authors assume that the observed counts for each pool are the allele counts of each marker in that pool, forcing all configuration matrices $\mathbf{A}_i$ to be identical. They also set the input haplotypes to be all $2^M$ haplotypes. \textcite{pirinen_estimating_2009} provides an implementation of this frequentist approach that instead allows the user to specify an arbitrary list of $H$ input haplotypes, known as `AEM algorithm with List' (AEML).

\textcite{pirinen_estimating_2009} introduced a Bayesian approach where the list of input haplotypes is treated as random, instead of being specified by the user. This is achieved by specifying a joint prior distribution over the number of input haplotypes, the configuration matrix, and the haplotype frequencies. The program HIPPO (\underline{H}aplotype estimation under \underline{i}ncomplete \underline{p}rior information
using \underline{p}ooled \underline{o}bservations) implements a reversible-jump MCMC sampler to perform inference on this model. Similar to \textcite{kuk_computationally_2009}, HIPPO also uses a normal approximation, and assumes that the observed counts for each pool are the allele counts of each marker in that pool. Since the list of input haplotypes is random, the configuration matrices $\mathbf{A}_i$ may not include all $2^M$ possible haplotypes, but are still identical across $i=1,\ldots,N$.

The accuracy of AEML and HIPPO hinges on the quality of the normal approximation. The exact marginal distribution of each observed count is a binomial distribution. Recall that the normal approximation to the binomial distribution $\mathrm{Bin}(n,p)$ is only accurate for sufficiently large $np(1-p)$. In the context of haplotype frequency estimation, some input haplotypes may be rare or even absent from the population. This leads to inaccuracies in the normal approximation for the case where some entries of $\mathbf{p}$ are small. HIPPO may suffer less from this issue as it is able to remove such input haplotypes from the configuration matrix during sampling. Moreover, numerical issues may arise when the covariance matrix in \eqref{eq:mn_approx} is nearly singular, which can happen if a pair of markers are highly correlated (high linkage disequilibrium), e.g. two markers where major alleles occur primarily together. We attempt to alleviate numerical issues by adding a small stabilising constant to the diagonal, i.e. replacing the covariance matrix in \eqref{eq:mn_approx} with ${n_i[\mathbf{A}_i(\mathrm{diag}(\mathbf{p})-\mathbf{p}\mathbf{p}^T)\mathbf{A}_i^T+\epsilon\,\mathbf{I}]}$, where $\epsilon=10^{-9}$ and $\mathbf{I}$ is the identity matrix. Nevertheless, near-singularity may still degrade the quality of the normal approximation, which we illustrate in Section~\ref{sec:approx_accuracy}.

\subsection{Proposed methods}\label{sec:proposed}

In this paper, we propose MCMC methods to perform Bayesian inference on the haplotype frequencies $\mathbf{p}$. We assume a Dirichlet prior with equal concentration $\alpha$ for the haplotype frequencies, i.e.
\begin{equation}
\mathbf{p}\sim\mathrm{Dir}(\alpha,\ldots,\alpha)\label{eq:dir_prior}.
\end{equation}
Unlike AEML and HIPPO, we relax the assumption that observed counts are allele counts for our methods. A motivating example can be found from genetic studies on sulfadoxine-pyrimethamine (SP) resistance in \textit{Plasmodium falciparum} parasites. There are primarily 3 SNPs of interest on the \textit{dhps} gene that are indicative of SP resistance, namely \textit{dhps}437/540/581~\parencite{sibley_sp}. However, some studies only report haplotype data over 2 markers, \textit{dhps}437 and \textit{dhps}540. This can be understood as the observed counts being the counts of 4 partial haplotypes (over the 2 markers). Each partial haplotype in turn corresponds to a subset of the full haplotypes (over all 3 markers); in this case each subset consists of 2 full haplotypes, as we consider \textit{dhps}581 to have two possible alleles. Our methods include the flexibility for each observed count to correspond to a different subset of the full haplotypes, which is not implemented in AEML and HIPPO. We assume that the user specifies the $H$ input haplotypes, and determines the configuration matrices $\mathbf{A}_i$ from the nature of the observed counts.

All of our methods require some preprocessing of the configuration matrices. For each $i=1,\ldots,N$, we include the pool size $n_i$ as an entry of the observed count vector $\mathbf{y}_i$, where the corresponding row in $\mathbf{A}_i$ is a row of 1s. This absorbs the equality condition \eqref{eq:sum_to_n} into the linear system \eqref{eq:linear_sys}. If any configuration matrix $\mathbf{A}_i$ is not of full row rank, we use row reduction to obtain a submatrix consisting of a maximal set of linearly independent rows. This removes redundant information observed from a latent multinomial model, see \textcite{zhang_fast_2019} for an explanation of why inference results are not affected by this procedure. Hereafter, we assume that all configuration matrices are of full row rank.

\subsubsection{Marginalisation}

Our first approach is to marginalise out the latent counts $\mathbf{Z}\coloneqq\{\mathbf{z}_i\}_{i=1}^n$ from the likelihood
\begin{equation}
p( \mathbf{Y},\mathbf{Z} \vert \mathbf{p} ) = \prod_{i=1}^N p(\mathbf{y}_i, \mathbf{z}_i\vert \mathbf{p}) = \prod_{i=1}^N \underbrace{ \binom{n_i}{z_{i1},\ldots,z_{iH}} p_1^{z_{i1}}\cdots p_H^{z_{iH}} }_{ p(\mathbf{z}_i \vert \mathbf{p}) } \underbrace{\vphantom{\binom{n_i}{z_{i1},\ldots,z_{iH}}} \mathds{1}(\mathbf{y}_i=\mathbf{A}_i\mathbf{z}_i)}_{ p(\mathbf{y}_i \vert \mathbf{z}_i) } ,
\end{equation}
where $\mathbf{Y}\coloneqq\{\mathbf{y}_i\}_{i=1}^n$. This amounts to computing $p(\mathbf{y}_i\vert \mathbf{p})$ by enumerating all possible latent counts $\mathbf{z}_i$ for each $i=1,\ldots,N$. Although this approach does not scale well with pool size, it is still appropriate for cases where the number of input haplotypes is small enough. For each $i=1,\ldots,N$, we define the \emph{feasible set} to be the set of solutions to \eqref{eq:linear_sys}--\eqref{eq:nonneg}, i.e.
\begin{equation*}
    \mathcal{F}(\mathbf{A}_i,\mathbf{y}_i) \coloneqq
    \{ \mathbf{z}_i \colon \mathbf{A}_i\mathbf{z}_i=\mathbf{y}_i, z_{i1}\ge 0, \ldots, z_{iH}\ge 0 \},
\end{equation*}
where the equality condition~\eqref{eq:sum_to_n} is absorbed into the linear system $\mathbf{A}_i\mathbf{z}_i=\mathbf{y}_i$. The probability mass function of $\mathbf{y}_i$ is given by
\begin{equation}
    p(\mathbf{y}_i\vert\mathbf{p}) = \, \smashoperator[l]{\sum_{\mathbf{z}_i\in\mathcal{F}(\mathbf{A}_i,\mathbf{y}_i)}} \! p(\mathbf{z}_i\vert\mathbf{p}) = \, \smashoperator[l]{\sum_{\mathbf{z}_i\in\mathcal{F}(\mathbf{A}_i,\mathbf{y}_i)}} \! \binom{n_i}{z_{i1},\ldots,z_{iH}} p_1^{z_{i1}}\cdots p_H^{z_{iH}}. \label{eq:exact}
    \vspace{-5pt}
\end{equation}
To perform Bayesian inference, we first enumerate the feasible sets $\mathcal{F}(\mathbf{A}_i,\mathbf{y}_i)$ for each $i=1,\ldots,N$, then proceed with running MCMC to obtain samples from the posterior distribution $p(\mathbf{p}\vert\mathbf{Y})$. We perform MCMC using the No-U-Turn sampler (NUTS)~\parencite{hoffman_no-u-turn_2011} as implemented in \texttt{PyMC}~\citep{Salvatier2016}. NUTS simulates a Markov chain that converges to the posterior distribution by utilising gradient information of the log-posterior, which avoids the inefficient random-walk behaviour exhibited by traditional Metropolis-Hastings proposals. 

In the case where a configuration matrix $\mathbf{A}_i$ consists of arbitrary integer entries, one can enumerate the feasible set with \texttt{4ti2}~\parencite{4ti2}, a software package for `algebraic, geometric and combinatorial problems on linear spaces'. However, our configuration matrices only have 0s and 1s as entries, which allows for a more efficient branch-and-bound algorithm for finding the feasible set as described in Appendix~\ref{sec:bnb}. If the number of input haplotypes or pool size is too large, the feasible set may have too many elements to be enumerated within a reasonable amount of time. In this case, we either resort to a normal approximation~\eqref{eq:mn_approx}, or sample the latent counts instead of marginalising them out, as described in the next section.

\subsubsection{Latent count sampling}
In order to avoid using approximations when the feasible set is too large, we treat latent counts $\mathbf{z}_i$ as model parameters to be sampled during MCMC alongside with $\mathbf{p}$. Sampling the latent counts $\mathbf{z}_i$ is not straightforward as the proposed values must belong to the feasible set. \textcite{gasbarra_estimating_2011} addresses this constraint by relaxing $\mathbf{z}_i$ to be continuous, and expressing each $\mathbf{z}_i$ as a convex combination of the extremal points of $\mathcal{F}(\mathbf{A}_i,\mathbf{y}_i)$. This approach comes at the cost of approximating the discrete multinomial distribution in \eqref{eq:multinom} with a continuous Dirichlet distribution. In this paper, we instead aim to sample discrete latent counts $\mathbf{z}_i$ using a custom Metropolis-within-Gibbs sampler without resorting to any approximations. Note that despite the connection between our approach and that of \textcite{gasbarra_estimating_2011}, we do not include their approach in our comparison as there is no software publicly available, and HIPPO has been shown to give better performance~\parencite{pirinen_estimating_2009}.

Before we describe our sampler, we first exploit the Dirichlet-multinomial conjugacy due to \eqref{eq:multinom} and \eqref{eq:dir_prior}. Define $z_{\boldsymbol{\cdot} h}\coloneqq z_{1h}+\cdots+z_{Nh}$ for each $h=1,\ldots,H$. The full conditional distribution of $\mathbf{p}$ is given by
\begin{align*}
p(\mathbf{p}\vert\mathbf{Y},\mathbf{Z}) 
&\propto p(\mathbf{p},\mathbf{Y},\mathbf{Z}) \\
&= \underbrace{\vphantom{\left[ \prod_{i=1}^N 
\binom{n_i}{z_{i1},\ldots,z_{iH}} \right]}
\frac{\Gamma(H\alpha)}{\Gamma(\alpha)^H}p_1^{\alpha-1}\cdots p_H^{\alpha-1}}_{p(\mathbf{p})} \underbrace{\left[ \prod_{i=1}^N 
\binom{n_i}{z_{i1},\ldots,z_{iH}} p_1^{z_{i1}}\cdots p_H^{z_{iH}}  \mathds{1}(\mathbf{A}_i\mathbf{z}_i=\mathbf{y}_i)\right]}_{p( \mathbf{Y},\mathbf{Z} \vert \mathbf{p} )} \\
&\propto p_1^{\alpha+z_{\boldsymbol{\cdot} 1}-1}\cdots p_H^{\alpha+z_{\boldsymbol{\cdot} H}-1},
\end{align*}
i.e. 
\begin{equation}
    \mathbf{p}\vert\mathbf{Y},\mathbf{Z} \sim \mathrm{Dir}(\alpha+z_{\boldsymbol{\cdot} 1},\ldots,\alpha+z_{\boldsymbol{\cdot} H}). \label{eq:dir_post}
\end{equation}
Moreover, we can marginalise $\mathbf{p}$ out from the joint distribution $p(\mathbf{p},\mathbf{Y},\mathbf{Z})$:
\begin{align}
    p(\mathbf{Y},\mathbf{Z})
    &= \int p(\mathbf{p},\mathbf{Y},\mathbf{Z}) \, d\mathbf{p} \nonumber \\
    &= \frac{\Gamma(H\alpha)}{\Gamma(\alpha)^H} \left[ \prod_{i=1}^N 
    \binom{n_i}{z_{i1},\ldots,z_{iH}}  \mathds{1}(\mathbf{A}_i\mathbf{z}_i=\mathbf{y}_i)\right]\int p_1^{\alpha+z_{\boldsymbol{\cdot} 1}-1}\cdots p_H^{\alpha+z_{\boldsymbol{\cdot} H}-1} \, d\mathbf{p} \nonumber \\
    &= \frac{\Gamma(H\alpha)}{\Gamma(\alpha)^H\Gamma(H\alpha+\sum_{i=1}^N n_i)}
    \left[ \prod_{i=1}^N \binom{n_i}{z_{i1},\ldots,z_{iH}}  \mathds{1}(\mathbf{A}_i\mathbf{z}_i=\mathbf{y}_i) \right] \prod_{h=1}^H \Gamma(z_{\boldsymbol{\cdot}h}+\alpha),  \label{eq:pyz}
\end{align}
where the integral is the normalising constant of a Dirichlet distribution.
This allows us to simulate posterior samples of $(\mathbf{Z},\mathbf{p})$ in two stages. We first obtain $S$ samples $\{\mathbf{Z}^{(s)}\}_{s=1}^S$ from $p(\mathbf{Z}\vert\mathbf{Y})$ using MCMC, which is possible as the unnormalised posterior $p(\mathbf{Y},\mathbf{Z})$ is available through~\eqref{eq:pyz}. For each MCMC sample $\mathbf{Z}^{(s)}$ where $s=1,\ldots,S$, we then sample $\mathbf{p}^{(s)}$ from $p(\mathbf{p}\vert\mathbf{Y},\mathbf{Z}=\mathbf{Z}^{(s)})$ using~\eqref{eq:dir_post}. The Markov chain $\{(\mathbf{Z}^{(s)},\mathbf{p}^{(s)})\}_{s=1}^S$ converges to the joint posterior $p(\mathbf{Z},\mathbf{p}\vert\mathbf{Y})$ since 
\begin{equation*}
p(\mathbf{Z},\mathbf{p}\vert\mathbf{Y}) = p(\mathbf{p}\vert\mathbf{Y},\mathbf{Z}) p(\mathbf{Z}\vert\mathbf{Y}).
\end{equation*}

For the remainder of this section, we describe a Metropolis-within-Gibbs (MwG) sampler for obtaining the samples $\{\mathbf{Z}^{(s)}\}_{s=1}^S$ from the posterior $p(\mathbf{Z}\vert\mathbf{Y})$. Let $\mathbf{Z}_{-i}\coloneqq\{ \mathbf{z}_1, \ldots, \mathbf{z}_{i-1}, \mathbf{z}_{i+1}, \ldots, \mathbf{z}_N \}$ for each $i=1,\ldots,N$. To specify the MwG sampler, we need to specify for each $i=1,\ldots,N$ a Metropolis-Hastings sampler whose target distribution is $p(\mathbf{z}_i\vert \mathbf{Y},\mathbf{Z}_{-i})$. Let $\mathbf{z}_i'$ denote the current value of $\mathbf{z}_i$ at any point of the sampler. In order to satisfy the constraint \eqref{eq:linear_sys}, we consider proposals that add or subtract a vector $\mathbf{u}$ chosen randomly from a subset $\mathcal{B}_i$ of the kernel of $\mathbf{A}_i$. Given that the current value $\mathbf{z}_i=\mathbf{z}_i'$ satisfies \eqref{eq:linear_sys}, the resulting proposal $\mathbf{z}_i = \mathbf{z}_i' \pm \mathbf{u}$ will also satisfy \eqref{eq:linear_sys}. \textcite{link_uncovering_2010} set the subset $\mathcal{B}_i$ to be an arbitrary basis of $\mathbf{A}_i$, however, the resulting Markov chain may not be irreducible~\parencite{schofield_connecting_2015}, i.e. some points in $\mathcal{F}(\mathbf{A}_i,\mathbf{y}_i)$ may never be reached. This is because if the only `moves' are vectors of an arbitrary basis, there may be points of the feasible set that can only be reached through points with negative entries, which violates \eqref{eq:nonneg}. An alternative is to generate a proposal by adding linear combinations of the basis vectors to $\mathbf{z}_i'$. \textcite{diaconis_algebraic_1998} found this approach to be inefficient, as it generates proposals with negative entries too often. Instead, the authors proposed to use a larger subset $\mathcal{B}_i$ of the kernel of $\mathbf{A}_i$, such that all points of the feasible set may be reached through points with nonnegative entries only. Such a subset $\mathcal{B}_i$ is known as a \emph{Markov basis} of $\mathbf{A}_i$, and satisfies the condition that a graph with $\mathcal{F}(\mathbf{A}_i,\mathbf{y}_i)$ as its vertices and
\begin{equation*}
    \{ (\mathbf{v},\mathbf{w}) \colon \mathbf{v},\mathbf{w}\in\mathcal{F}(\mathbf{A}_i,\mathbf{y}_i),  \mathbf{v}-\mathbf{w}\in\mathcal{B}_i \text{ or } \mathbf{w}-\mathbf{v}\in\mathcal{B}_i \}
\end{equation*}
as its edges is always a connected graph for any vector $\mathbf{y}_i$ of $R_i$ nonnegative integers. The authors use techniques in commutative algebra to find the Markov basis of a matrix, which is implemented in \texttt{4ti2}~\parencite{4ti2}. 

Given a Markov basis $\mathcal{B}_i$ and the current value $\mathbf{z}_i = \mathbf{z}_i'$, we generate the proposal $\mathbf{z}_i^* = \mathbf{z}_i' + \delta\mathbf{u}$ with probability $q(\mathbf{z}_i^*\vert \mathbf{z}_i')$ proportional to $p(\mathbf{z}_i=\mathbf{z}_i^*\vert \mathbf{Y},\mathbf{Z}_{-i})$, where $\delta\in\{-1,1\}$ and $\mathbf{u}\in\mathcal{B}_i$. In other words, the proposal distribution is
\begin{align}
    q(\mathbf{z}_i^*\vert \mathbf{z}_i') &= \frac{p(\mathbf{z}_i=\mathbf{z}_i^*\vert \mathbf{Y},\mathbf{Z}_{-i})}{\sum_{\delta\in\{-1,1\}}\sum_{\mathbf{u}\in\mathcal{B}_i} p(\mathbf{z}_i=\mathbf{z}_i'+\delta\mathbf{u}\vert \mathbf{Y},\mathbf{Z}_{-i})}\nonumber\\
    &= \frac{p(\mathbf{Y},\mathbf{Z}_{-i},\mathbf{z}_i=\mathbf{z}_i^*)}{\sum_{\delta\in\{-1,1\}}\sum_{\mathbf{u}\in\mathcal{B}_i} p(\mathbf{Y},\mathbf{Z}_{-i},\mathbf{z}_i=\mathbf{z}_i'+\delta\mathbf{u})},\label{eq:mbasis_prop_conj}
\end{align}
where the formula for $p(\mathbf{Y},\mathbf{Z}_{-i},\mathbf{z}_i)$ is given in \eqref{eq:pyz}. Note that $p(\mathbf{Y},\mathbf{Z}_{-i},\mathbf{z}_i)$ is zero whenever $\mathbf{z}_i$ contains negative entries. The last equality in \eqref{eq:mbasis_prop_conj} follows from the fact that $p(\mathbf{z}_i\vert \mathbf{Y},\mathbf{Z}_{-i})$ is proportional to $p(\mathbf{Y},\mathbf{Z})$ as a function of $\mathbf{z}_i$. This proportionality also allows us to write the Metropolis-Hastings acceptance ratio as
\begin{align}
    a(\mathbf{z}_i^*; \mathbf{z}_i') &\coloneqq \min \left\{1, \frac{p(\mathbf{z}_i=\mathbf{z}_i^*\vert \mathbf{Y},\mathbf{Z}_{-i})}{p(\mathbf{z}_i=\mathbf{z}_i'\vert \mathbf{Y},\mathbf{Z}_{-i})} 
\frac{q(\mathbf{z}_i'\vert \mathbf{z}_i^*)}{q(\mathbf{z}_i^*\vert \mathbf{z}_i')} \right\} \nonumber \\
    &= \min \left\{1, \frac{\sum_{\delta\in\{-1,1\}}\sum_{\mathbf{u}\in\mathcal{B}_i} p(\mathbf{Y},\mathbf{Z}_{-i},\mathbf{z}_i=\mathbf{z}_i'+\delta\mathbf{u})}{\sum_{\delta\in\{-1,1\}}\sum_{\mathbf{u}\in\mathcal{B}_i} p(\mathbf{Y},\mathbf{Z}_{-i},\mathbf{z}_i=\mathbf{z}_i^*+\delta\mathbf{u})} \right\}.\label{eq:acc_rate_conj}
\end{align}
The choice of a proposal distribution~\eqref{eq:mbasis_prop} that is proportional to the full conditional distribution can be considered as a restricted Gibbs proposal, though the entire support of $\mathbf{z}_i$ is unlikely to be covered by one proposal iteration. Nevertheless, the use of a Markov basis guarantees the chain to be irreducible. Note that the proposal distribution~\eqref{eq:mbasis_prop} is different from that of \textcite{schofield_connecting_2015}, who sample the basis vector $\mathbf{u}$ uniformly. \citet{hazelton_dmb} show that a Gibbs-like proposal explores the posterior distribution more efficiently due to a higher acceptance rate.

\begin{algorithm*}[t]
\caption{\small Collapsed random-scan Metropolis-within-Gibbs sampler for the latent multinomial model with Dirichlet conjugacy. $T$ is the number of burn-in iterations, $S$ is the number of inference iterations, $C$ is the number of latent count updates per iteration.}\label{algo:mwg_dirmult}
\DontPrintSemicolon
\small
\KwIn{Initial values $\{\mathbf{z}_i^{(0)}\}_{i=1}^N$, Markov bases $\{\mathcal{B}_i\}_{i=1}^N$}
\KwOut{Posterior samples $\{ \mathbf{p}^{(s)}, \mathbf{Z}^{(s)}\}_{s=1}^{S}$}
\For{$i\gets 1$ \KwTo $N$}{
        $\mathbf{z}'_i \gets \mathbf{z}_i^{(0)}$\;
    }
\For{$t\gets 1$ \KwTo $T+S$}{
    \For{$c\gets 1$ \KwTo $C$}{
        randomly select $i$ from $\{ 1,\ldots,N\}$ with probability proportional to $n_i$ \;
        sample $\mathbf{z}_i^* = \mathbf{z}_i'+\delta\mathbf{u}$ according to $q(\mathbf{z}_i^*\vert\mathbf{z}'_i)$ from \eqref{eq:mbasis_prop_conj} \;
        replace $\mathbf{z}'_i$ with $\mathbf{z}^*_i$ with probability $a(\mathbf{z}^*_i;\mathbf{z}'_i)$ from \eqref{eq:acc_rate_conj}\;
    }
    \If{$t > T$}{
        $s \gets t - T$ \;
        $(\mathbf{z}_1^{(s)}, \ldots, \mathbf{z}_N^{(s)}) \gets (\mathbf{z}'_1, \ldots, \mathbf{z}'_N)$\;
        \For{$h\gets 1$ \KwTo $H$}{
            $\mathbf{z}_{\boldsymbol{\cdot} h}^{(s)} \gets z^{(s)}_{1h}+\cdots+z^{(s)}_{Nh}$\;
        }
        sample $\mathbf{p}^{(s)} \sim \mathrm{Dir}\!\left(\alpha+z^{(s)}_{\boldsymbol{\cdot} 1},\ldots,\alpha+z^{(s)}_{\boldsymbol{\cdot} H}\right)$ according to \eqref{eq:dir_post}\;
    }
}
\KwRet $\{\mathbf{p}^{(s)},\mathbf{Z}^{(s)}\}_{s=1}^{S}$\;
\end{algorithm*}

Augmenting the MwG sampler for $p(\mathbf{Z}\vert\mathbf{Y})$ with sampling $\mathbf{p}$ according to \eqref{eq:dir_post} leads to a collapsed MwG sampler~\parencite{liu_collapsed_1994}, which we describe in Algorithm~\ref{algo:mwg_dirmult}. The sampler starts with $T$ burn-in iterations, where the samples are discarded as the chain may not have converged to the posterior distribution. We use a random scan order when updating the latent counts, where the probability of choosing $\mathbf{z}_i$ to update is proportional to the pool size $n_i$ as the corresponding feasible set grows in size with $n_i$. We perform $C$ such updates every iteration, where larger values of $C$ lead to less autocorrelation in the posterior samples at the cost of longer computational runtime. We set $C$ to be proportional to the total pool size $n_1+\cdots+n_N$. The initial values for $\mathbf{Z}$ can be found by solving \eqref{eq:linear_sys}--\eqref{eq:nonneg} using integer programming methods.

\subsection{Partition ligation for determining input haplotypes}\label{sec:partition_ligation}

For a moderate number of markers ($M\ge 6$), the number of haplotypes present in a population is typically much smaller than the number of possible haplotypes, $2^M$. For our methods to be scalable with the number of markers, we need to prevent the number of input haplotypes from growing exponentially with $M$. This is not a concern if a complete list of the haplotypes present is available. If the list is incomplete, we use \emph{partition ligation}~\parencite{niu_bayesian_2002} to determine input haplotypes, i.e. haplotypes whose frequencies we will infer. We first segment the sequence of $M$ markers into blocks of 3 or 4 markers. We call the haplotypes implicated over a block of markers \emph{partial haplotypes}. The idea of partition ligation is to construct full input haplotypes 
by combining from each block the partial haplotypes with the highest estimated frequencies. First, we obtain point estimates of the frequencies of the partial haplotypes from each block using one of the methods from Section~\ref{sec:existing} or \ref{sec:proposed}. In this paper, we perform this using MCMC-Approx, and use the posterior mean as the point estimate. Suppose we have $b$ blocks $B_1, \ldots, B_b$ of markers. For $i=1,\ldots,b$, let $\mathcal{H}_i$ be the set of partial haplotypes from block $B_i$ whose point estimates are larger than some threshold $f$. For each $j=1,\ldots,\lfloor b/2\rfloor$, we concatenate every partial haplotype in $\mathcal{H}_{2j-1}$ with every partial haplotype in $\mathcal{H}_{2j}$ to form the set of haplotypes for the concatenated block $B_{2j-1}B_{2j}$. This procedure halves the numbers of blocks, and is repeated recursively until all blocks are concatenated together. The final list of concatenated haplotypes are used as the input haplotypes for subsequent inference. Choosing a lower threshold for $f$ makes it more likely for the constructed input haplotypes to include all haplotypes present in the population, but also introduces more input haplotypes that do not occur in the population, making subsequent inference less efficient. Details of partition ligation are further described in haplotype phasing literature, see for example, \textcite{stephens_comparison_2003}.

\subsection{Hierarchical extension}\label{sec:hier_ext}

\begin{figure}[t]
\centering
\includegraphics[width=0.7\textwidth]{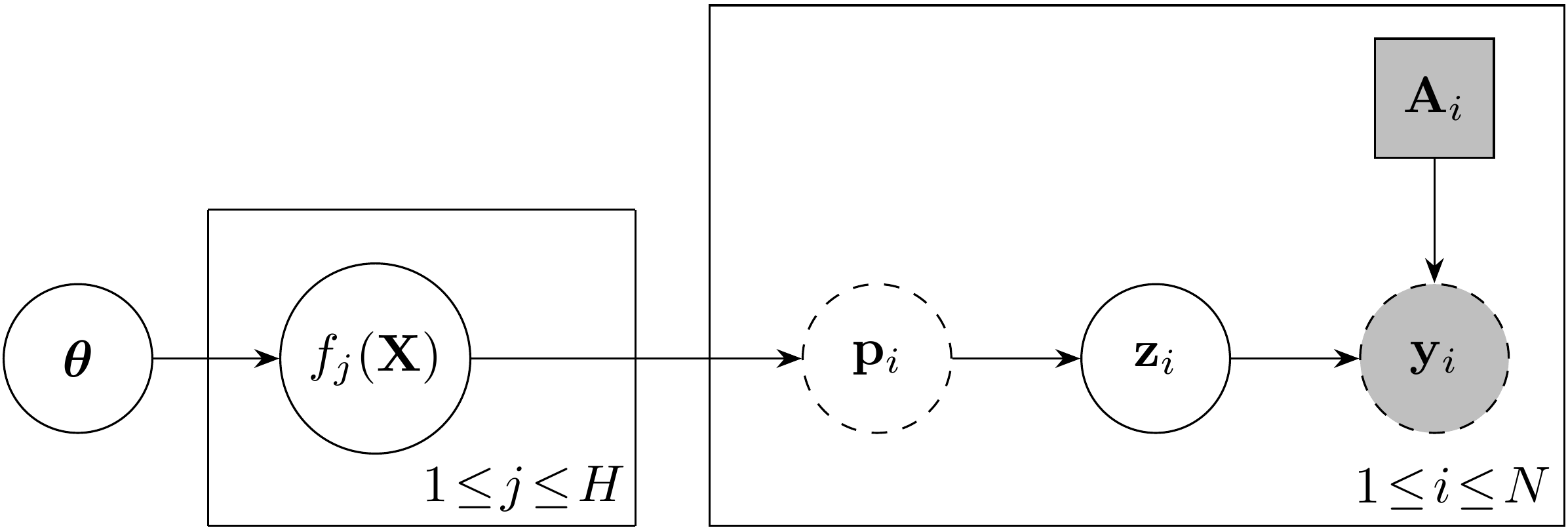}
\caption{\small Graphical model for latent multinomial data with multiple populations whose haplotype frequencies $\{\mathbf{p}_i\}_{i=1}^N$ are correlated through Gaussian processes. $f_h(\mathbf{X})$ denotes the vector $(f_h(\mathbf{x}_1),\ldots,f_h(\mathbf{x}_N))$. Circles and squares correspond to random variables and constants respectively. A shaded node indicates that the variable is observed. A dotted outline indicates that the variable is deterministically calculated from its parent variables. Variables contained within a plate are repeated according to the index at the bottom right.}\label{fig:hierarchical_graphical}
\end{figure}

In meta-analysis studies, genetic data collected from multiple populations are analysed together, where each population has its own set of haplotype frequencies. We extend the latent multinomial model \eqref{eq:multinom}--\eqref{eq:linear_sys} to a hierarchical model where each pool of samples is drawn from a different population. To account for the correlation between haplotype frequencies of different populations, we model the haplotype frequencies as a softmax transformation of $H$ Gaussian processes (GPs):
\begin{align}
\mathbf{y}_i &= \mathbf{A}_i\mathbf{z}_i & \text{for } i=1,\ldots,N, \\
\mathbf{z}_i \vert \mathbf{p}_i &\sim \mathrm{Mult}(n_i, \mathbf{p}_i) & \text{for } i=1,\ldots,N, \label{eq:multinom_hierarchical}\\
p_{ih} &= \frac{\exp(f_h(\mathbf{x}_i))}{\exp(f_1(\mathbf{x}_i))+\cdots+\exp(f_H(\mathbf{x}_i))} & \text{for } i=1,\ldots,N, \, h=1,\ldots,H, \\
f_h(\mathbf{x}_1), \ldots,  f_h(\mathbf{x}_N) &\sim \mathrm{N}(\mathbf{m}_h(\mathbf{X}), \mathbf{C}_h(\mathbf{X},\mathbf{X})) & \text{for } h=1,\ldots,H,
\end{align}
where $\mathbf{p}_i$ are the haplotype frequencies of population $i$, $\mathbf{X} = \{\mathbf{x}_i\}_{i=1}^N$ are the covariates observed for each population, and $f_h$ is the $h$-th GP whose mean function and covariance function are $\mathbf{m}_h$ (vector-valued) and $\mathbf{C}_h$ (matrix-valued) respectively. The mean and covariance functions are further parametrised by GP hyperparameters $\bm\theta$. A graphical representation of this model is shown in Figure~\ref{fig:hierarchical_graphical}.

As an example, we consider time-series modelling of haplotype frequencies, where the only covariate for each population $i$ is the time of data collection $t_i$. We specify each mean function to be a constant $\mathbf{m}_h(\mathbf{X}) = (\mu_h\ldots,\mu_h)^T$, and each covariance function to be the sum of a rational quadratic kernel and a white noise kernel, i.e. the $(i,i')$-th entry of $\mathbf{C}_h(\mathbf{X},\mathbf{X})$ is
\begin{equation}
c_h(t_i,t_{i'}) \coloneqq s_h^2 \! \left( 1 + \frac{(t_i - t_{i'})^2}{2\tau_h^2} \right)^{-1} \hspace{-0.5em} + \sigma^2 \mathds{1}(i=i'), \label{eq:kernel_example}
\end{equation}
where $\tau_h$ is the timescale, $s_h$ is the temporal standard deviation, $\sigma$ is the noise standard deviation, and $\mathds{1}(\cdot)$ is the indicator function. Pools that are observed closer in time have haplotype frequencies that are more strongly correlated since $c(t_i,t_{i'})$ increases as $\lvert t_i-t_{i'}\rvert$ decreases. The noise term $\sigma^2 \mathds{1}(i=i')$ accounts for overdispersion of the multinomial counts. The GP hyperparameters $\bm\theta \coloneqq (\{ \mu_h, \tau_h, s_h \}_{h=1}^H,\sigma)$ are given priors according to domain knowledge. 

Given the large number of continuous parameters, we perform MCMC inference with NUTS for the parameters $\mathbf{P} \coloneqq \{\mathbf{p}_i\}_{i=1}^N$ and $\bm\theta$. To deal with the latent counts, we may either use (i)~exact marginalisation by enumerating feasible sets, (ii)~approximate marginalisation according to \eqref{eq:mn_approx}, or (iii)~latent count sampling. It is straightforward to apply NUTS to both of the marginalisation approaches. The latent count sampling approach requires modification as the use of a GP prior implies that we no longer have Dirichlet-multinomial conjugacy. We instead use a MwG sampler with target distributions $p(\mathbf{P},\bm\theta \vert \mathbf{Z})$ and $p(\mathbf{z}_i \vert \mathbf{Y}, \mathbf{Z}_{-i}, \mathbf{P}, \bm\theta)$ for each $i=1,\ldots,N$.

\begin{algorithm*}[t]
\caption{\small Metropolis-within-Gibbs sampler for the latent multinomial model with a GP hierarchical extension. $T$ is the number of burn-in iterations, $S$ is the number of inference iterations, $C_i$ is the number of updates per iteration for $\mathbf{z}_i$.}\label{algo:mwg_gp}
\DontPrintSemicolon
\small
\KwIn{Initial values $\{\mathbf{z}_i^{(0)}\}_{i=1}^N$, Markov bases $\{\mathcal{B}_i\}_{i=1}^N$}
\KwOut{Posterior samples $\{ \mathbf{P}^{(s)}, \bm\theta^{(s)}, \mathbf{Z}^{(s)}\}_{s=1}^{S}$}
\For{$i\gets 1$ \KwTo $N$}{
        $\mathbf{z}'_i \gets \mathbf{z}_i^{(0)}$\;
    }
\For{$t\gets 1$ \KwTo $T+S$}{
    sample $(\mathbf{P}', \bm\theta')$ from $p(\mathbf{P}, \bm\theta \vert \mathbf{Z} = (\mathbf{z}'_1, \ldots, \mathbf{z}'_N))$ using NUTS \;
    \For{$i\gets 1$ \KwTo $N$}{
        \For{$c\gets 1$ \KwTo $C_i$}{
            sample $\mathbf{z}_i^* = \mathbf{z}_i'+\delta\mathbf{u}$ according to $q(\mathbf{z}_i^*\vert\mathbf{z}'_i)$ from \eqref{eq:mbasis_prop} \;
            replace $\mathbf{z}'_i$ with $\mathbf{z}^*_i$ with probability $a(\mathbf{z}^*_i;\mathbf{z}'_i)$ from \eqref{eq:acc_rate}\;
        }
    }
    \If{$t > T$}{
        $s \gets t - T$ \;
        $(\mathbf{P}^{(s)}, \bm\theta^{(s)}, (\mathbf{z}^{(s)}_1, \ldots, \mathbf{z}^{(s)}_N))\gets (\mathbf{P}', \bm\theta', (\mathbf{z}'_1, \ldots, \mathbf{z}'_N))$ \;
    }
}
\KwRet $\{\mathbf{P}^{(s)},\bm\theta^{(s)}, \mathbf{Z}^{(s)}\}_{s=1}^{S}$\;
\end{algorithm*}

Given pre-computed Markov bases $\mathcal{B}_i$ and the current value $\mathbf{z}_i=\mathbf{z}_i'$, we generate the proposal $\mathbf{z}_i^* = \mathbf{z}_i' + \delta\mathbf{u}$ with probability $q(\mathbf{z}_i^*\vert \mathbf{z}_i')$ proportional to $p(\mathbf{z}_i=\mathbf{z}_i^*\vert \mathbf{Y},\mathbf{Z}_{-i}, \mathbf{P}, \bm\theta)$, where $\delta\in\{-1,1\}$ and $\mathbf{u}\in\mathcal{B}_i$. We note that $p(\mathbf{z}_i\vert \mathbf{Y},\mathbf{Z}_{-i}, \mathbf{P}, \bm\theta)$ is proportional to $p(\mathbf{z}_i\vert \mathbf{p}_i)p(\mathbf{y}_i\vert \mathbf{z}_i)$, where $p(\mathbf{z}_i\vert \mathbf{p}_i)$ is given by \eqref{eq:multinom_hierarchical} and $p(\mathbf{y}_i\vert \mathbf{z}_i) = 1$ since any proposed value of $\mathbf{z}_i$ satisfies $\mathbf{A}_i\mathbf{z}_i = \mathbf{y}_i$. This allows us to write the proposal distribution as
\begin{align}
    q(\mathbf{z}_i^*\vert \mathbf{z}_i') &= \frac{p(\mathbf{z}_i=\mathbf{z}_i^*\vert \mathbf{Y},\mathbf{Z}_{-i}, \mathbf{P}, \bm\theta)}{\sum_{\delta\in\{-1,1\}}\sum_{\mathbf{u}\in\mathcal{B}_i} p(\mathbf{z}_i=\mathbf{z}_i'+\delta\mathbf{u}\vert \mathbf{Y},\mathbf{Z}_{-i}, \mathbf{P}, \bm\theta)}\nonumber\\
    &= \frac{p(\mathbf{z}_i=\mathbf{z}_i^* \vert \mathbf{p}_i)}{\sum_{\delta\in\{-1,1\}}\sum_{\mathbf{u}\in\mathcal{B}_i} p(\mathbf{z}_i=\mathbf{z}_i'+\delta\mathbf{u} \vert \mathbf{p}_i)} ,\label{eq:mbasis_prop}
\end{align}
and the acceptance ratio as
\begin{align}
    a(\mathbf{z}_i^*; \mathbf{z}_i') &= \min \left\{1, \frac{p(\mathbf{z}_i=\mathbf{z}_i^*\vert \mathbf{Y},\mathbf{Z}_{-i}, \mathbf{P}, \bm\theta)}{p(\mathbf{z}_i=\mathbf{z}_i'\vert \mathbf{Y},\mathbf{Z}_{-i}, \mathbf{P}, \bm\theta)} 
\frac{q(\mathbf{z}_i'\vert \mathbf{z}_i^*)}{q(\mathbf{z}_i^*\vert \mathbf{z}_i')} \right\} \nonumber \\
    &= \min \left\{1, \frac{\sum_{\delta\in\{-1,1\}}\sum_{\mathbf{u}\in\mathcal{B}_i} p(\mathbf{z}_i=\mathbf{z}_i'+\delta\mathbf{u}\vert \mathbf{p}_i)}{\sum_{\delta\in\{-1,1\}}\sum_{\mathbf{u}\in\mathcal{B}_i} p(\mathbf{z}_i=\mathbf{z}_i^*+\delta\mathbf{u}\vert \mathbf{p}_i)} \right\}.\label{eq:acc_rate}
\end{align}
As for the target distribution $p(\mathbf{P},\bm\theta \vert \mathbf{Z})$, we use NUTS to propose MCMC samples $\{\mathbf{P}^{(s)},\bm\theta^{(s)}\}_{s=1}^S$. The full MCMC scheme is described in Algorithm~\ref{algo:mwg_gp}. Since updating $\mathbf{z}_i$ only depends on its current value and $\mathbf{p}_i$, there is no need for a random scan order. The number of updates for $\mathbf{z}_i$ is denoted as $C_i$, which we set to be proportional to the pool size $n_i$.

\section{Results}\label{sec:results}


We implement three MCMC methods: `MCMC-Exact' marginalises out $\mathbf{Z}$ exactly using \eqref{eq:exact}, `MCMC-Approx' marginalises out $\mathbf{Z}$ approximately using \eqref{eq:mn_approx}, and `LC-Sampling' samples $\mathbf{Z}$ according to Algorithm~\ref{algo:mwg_dirmult} or Algorithm~\ref{algo:mwg_gp} depending on whether the haplotype frequencies are shared across pools. The code is available at \url{https://github.com/ysfoo/haplm}. We present four sets of results: (i) a comparison of the exact likelihood \eqref{eq:exact} and the approximate likelihood \eqref{eq:mn_approx} based on a toy example, (ii) a comparison of our methods and existing methods (AEML and HIPPO) based on synthetic data, (iii) a comparison of our methods and existing methods based on real human data, and (iv) a demonstration of our methods applied to time-series data in a hierarchical setting. For all examples, the observed data consists of the allele counts of each marker in each pool.

\subsection{Accuracy of normal approximation}\label{sec:approx_accuracy}

\begin{figure}[t]
\centering
\includegraphics[width=\textwidth]{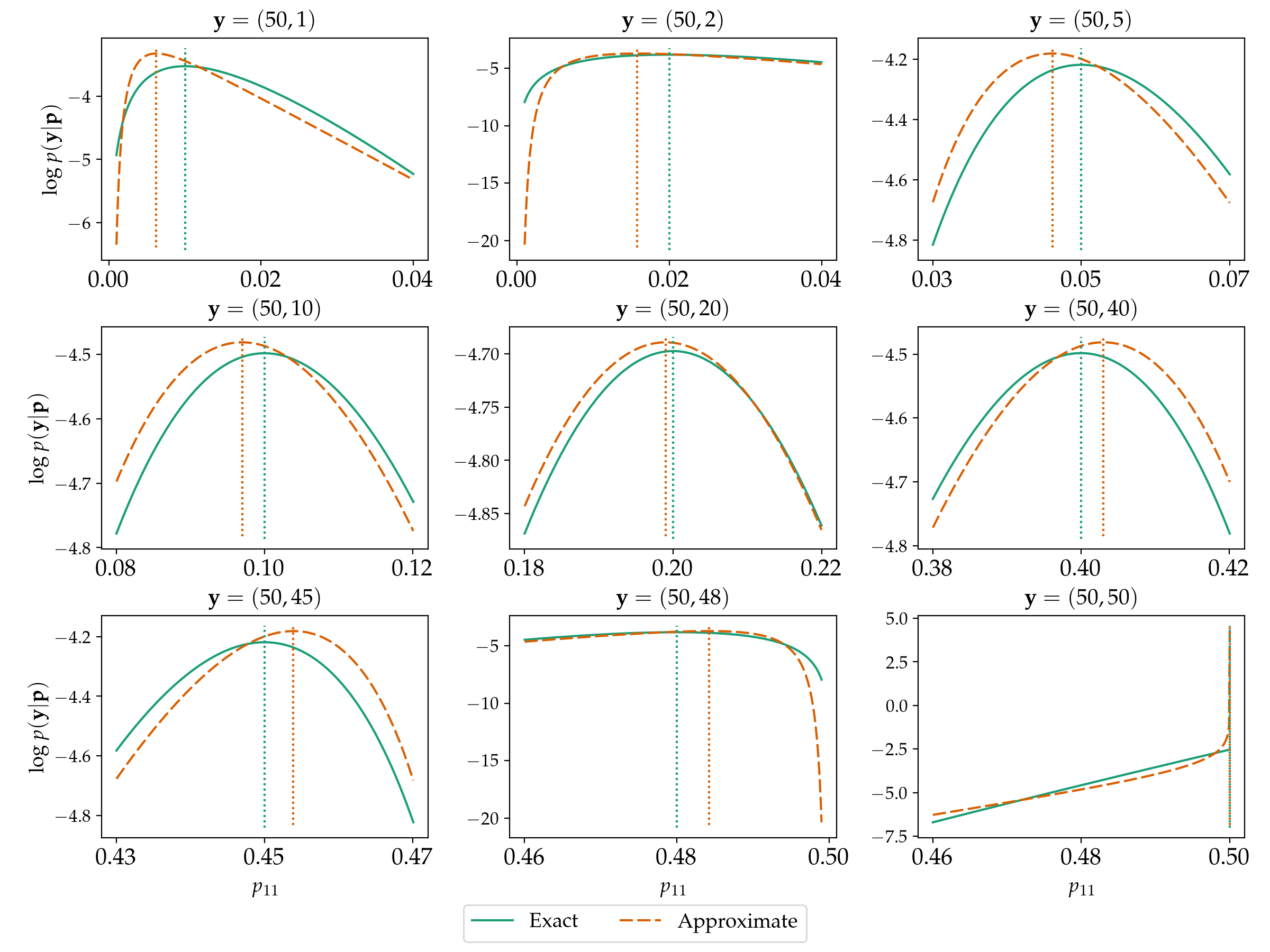}
\caption{\small Exact (solid) and approximate (dashed) log-likelihoods $p(\mathbf{y}\vert\mathbf{p})$ evaluated at haplotype frequencies ${\mathbf{p}=(0.5, 0.5-p_{11}, 0, p_{11})}$, where $\mathbf{y}$ consists of allele counts across two markers for one pool of size $n=100$. The dotted lines indicate where the exact and approximate log-likelihoods are maximised.}\label{fig:mn_vs_exact_mle}
\end{figure}

In this section, we illustrate cases where the normal approximation~\eqref{eq:mn_approx} is inaccurate, even when applied to a large pool of $100$ samples. Consider the simplest example where we have one data point $\mathbf{y}=(y_1,y_2)$ of allele counts across $M=2$ markers for a pool of $n$ haplotype samples. We denote the haplotype frequencies as $\mathbf{p} \coloneqq (p_{00}, p_{10}, p_{01}, p_{11})$, where $p_h$ is the frequency of haplotype $h$. We set the pool size to be $n=100$ and the allele count of the first marker to be $y_1=50$, and vary $y_2$ between 1 and 50. We find that the exact likelihood~\eqref{eq:exact} is maximised for two sets of haplotype frequencies: $\hat{\mathbf{p}} = (0.5, 0.5-y_2/n, 0, y_2/n)$ and $\hat{\mathbf{p}}' = (0.5-y_2/n, 0.5, y_2/n, 0)$, i.e. these are the exact maximum likelihood estimators (MLEs). In Figure~\ref{fig:mn_vs_exact_mle}, we compare the exact likelihood~\eqref{eq:exact} and the approximate likelihood~\eqref{eq:mn_approx} for values of $\mathbf{p}$ that are close to the first MLE, $\hat{\mathbf{p}}$, for various values of $y_2$. Since $y_2$ has no effect on the entries $(p_{00}, p_{01})$ of the first MLE, we only vary the values of $(p_{10}, p_{11})$ in our comparison. Overall, the values of $(p_{10}, p_{11})$ that maximise the exact and approximate likelihoods do not differ by more than 0.01. However, we notice that the normal approximation is less accurate when $y_2$ is close to 0 or 50. In fact, the approximate likelihood increases without bound as $\mathbf{p}\rightarrow (0.5,0,0,0.5)$ when $y_2=50$, while the exact likelihood remains bounded. This is because the covariance matrix in \eqref{eq:mn_approx} becomes singular as $\mathbf{p}\rightarrow (0.5,0,0,0.5)$. In general, the covariance matrix may become singular when certain entries of $\mathbf{p}$ approach zero. As such, the accuracy of the normal approximation depends on the data observed: if the data observed supports values of $\mathbf{p}$ such that the covariance matrix becomes near-singular, then the frequency of rare haplotypes may be underestimated.

\subsection{Synthetic data with shared haplotype frequencies}\label{sec:sim_shared}

\begin{figure}[t]
\centering
\includegraphics[width=\textwidth]{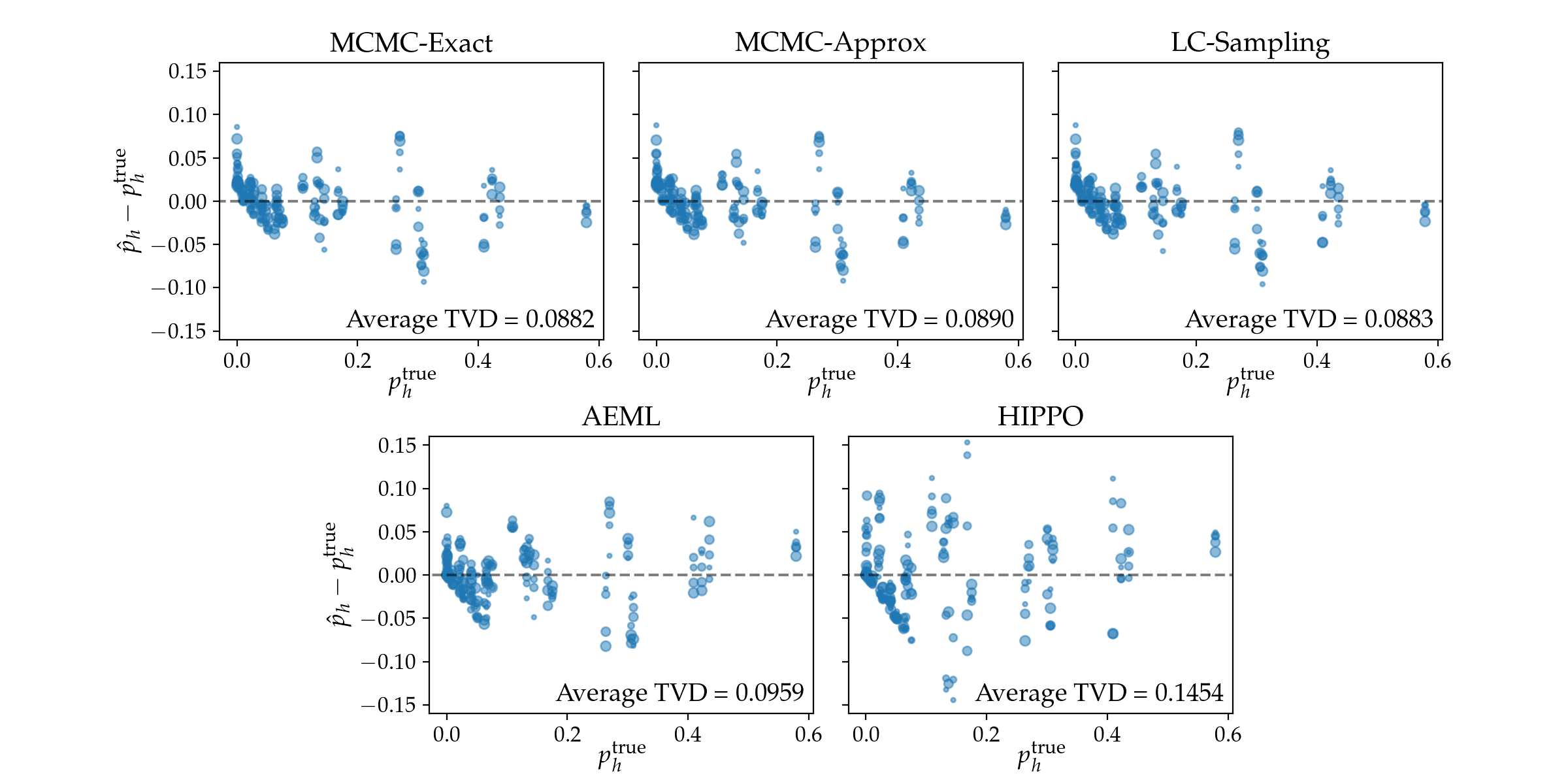}
\caption{\small Statistical performance of point estimates $\hat{\mathbf{p}}$ across 25 synthetic datasets where pools share the same true haplotype frequencies, $\mathbf{p}^\text{true}$. The errors $\hat{p}_h - p^\text{true}_h$ are plotted against each true haplotype frequency $p^\text{true}_h$. The size of each point is scaled by the pool size, $N$. The average (over 25 datasets) TVD between true haplotype frequencies and point estimates is shown in the bottom right of each plot.} \label{fig:synthetic_shared_errors}
\end{figure}

\begin{figure}[t]
\centering
\includegraphics[width=\textwidth]{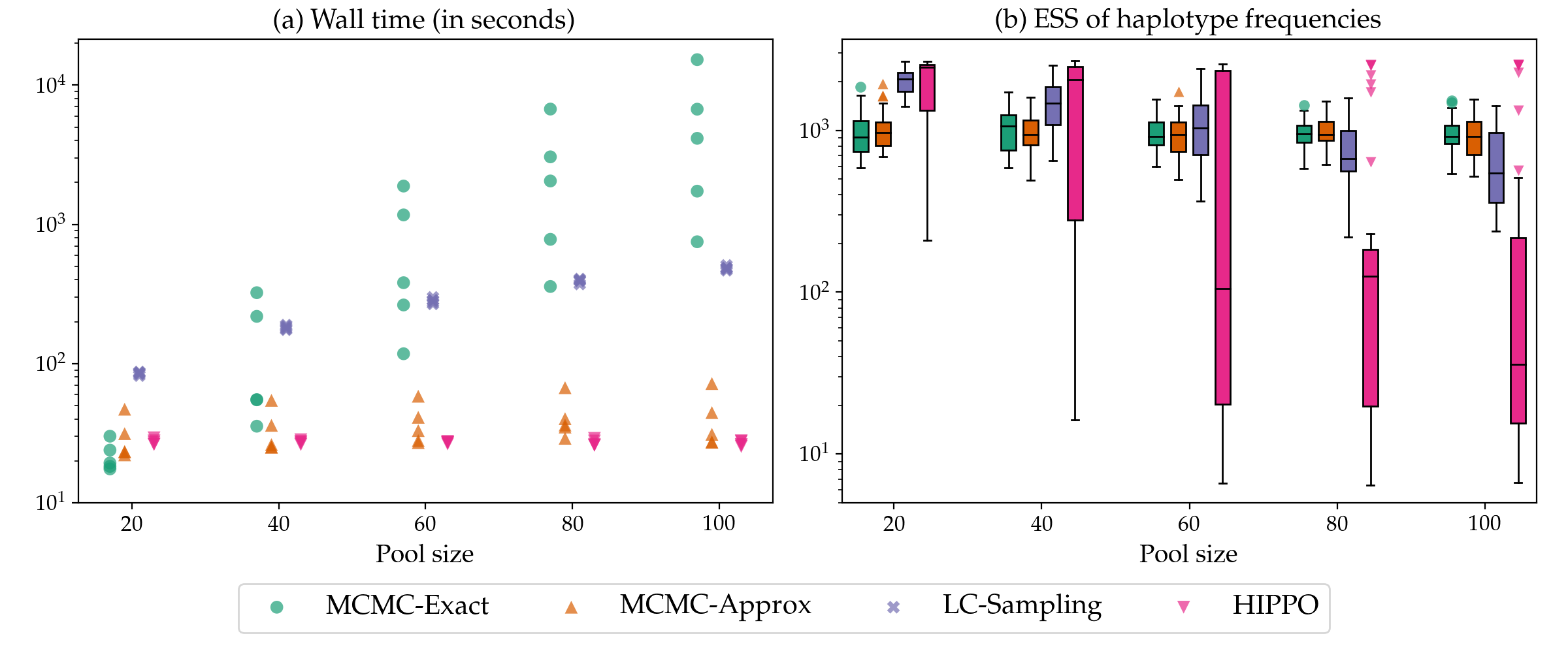}
\caption{\small (a)~Computational wall times and (b)~boxplots of ESS for haplotype frequencies $\{p_h\}_{1\le h\le H}$, across all datasets against the number of samples per pool for Bayesian methods applied to 25 synthetic datasets. Each boxplot corresponds to the haplotype frequencies over 5 datasets with the same pool size.} \label{fig:synthetic_shared_comp}
\end{figure}

To evaluate our three proposed methods, we first compare their statistical and computational performance with AEML and HIPPO when applied to synthetic datasets where all pools within a dataset share the same haplotype frequencies. We use the default parameters and settings when running AEML and HIPPO according to programs provided by \textcite{pirinen_estimating_2009}. For all MCMC methods, we run 5 chains for each method. Different MCMC methods require different chain lengths to reach convergence. For this example, having 500 burn-in iterations and 500 inference iterations per chain is sufficient for our proposed methods, as NUTS uses gradient information of the posterior to produce chains with low autocorrelation. On the other hand, \textcite{pirinen_estimating_2009} recommends $5\times 10^5$ iterations per chain for HIPPO as it produces chains with higher autocorrelation. We report the \emph{effective sample size} (ESS), which estimates the equivalent number of independent samples such that the information provided by that many independent samples is equivalent to that of the MCMC samples. In order to compare ESS across methods, we thin each chain to 500 samples per chain, regardless of the MCMC method that produced it. For LC-Sampling, the parameter $C$ from Algorithms~\ref{algo:mwg_dirmult} acts as a thinning factor, which we set to $C=5(n_1+\cdots+n_N)$.

We simulate 5 sets of haplotype frequencies $\mathbf{p}^\text{true}$ over $M=3$ markers from the distribution $\textrm{Dir}(0.4,\ldots,0.4)$, which induces some sparsity in $\mathbf{p}^\text{true}$. For each $\mathbf{p}^\text{true}$, we in turn simulate 5 datasets (each with $N=20$ pools) where the pool size is set to $n=20,40,60,80,100$, giving a total of 25 datasets. Latent haplotype counts are sampled according to the frequencies $\mathbf{p}^\text{true}$. The number of distinct haplotypes in each of our simulated datasets range between 6 and 8. All $H=8$ possible haplotypes are used as our input haplotypes. 

We compare the following point estimates: the posterior means under MCMC-Exact, MCMC-Approx, LC-Sampling, HIPPO, and the MLE under AEML. We measure the distance between a point estimate $\hat{\mathbf{p}}$ and the true frequencies $\mathbf{p}^\text{true}$ by the \emph{total variation distance} (TVD):
\begin{equation}
    \mathrm{TVD}(\hat{\mathbf{p}}, \mathbf{p}^\text{true}) \coloneqq \frac{1}{2} \sum_{h=1}^{2^M} \lvert \hat{p}_h - p^\text{true}_h \rvert. \label{eq:tvd}
\end{equation}
TVD can be interpreted as the probability mass redistributed to turn one haplotype distribution into another. In general, the summation in \eqref{eq:tvd} is taken not only over the input haplotypes but all possible haplotypes, as the true distribution may include haplotypes absent from the input haplotypes, e.g. when partition ligation (Section~\ref{sec:partition_ligation}) is used to determine the input haplotypes.

In Figure~\ref{fig:synthetic_shared_errors}, we report the TVDs between the true frequencies and each point estimate, and plot the errors $\hat{p}_h - p^\text{true}_h$ for each haplotype $h$ against the true haplotype frequencies. The results for our proposed methods (top row) are very similar. There is a diagonal on the left end of all plots, corresponding to $\hat{p}_h\approx 0.02$ for our three proposed methods, and $\hat{p}_h\approx 0$ for AEML and HIPPO. The average TVDs under AEML and HIPPO are larger, indicating less accurate inference. As seen in Section~\ref{sec:approx_accuracy}, the approximate likelihood can become unbounded when some haplotype frequencies are zero, which may explain the diagonal around $\hat{p}_h\approx 0$ for the maximum likelihood method AEML. On the other hand, HIPPO may remove rare haplotypes from the list of input haplotypes during MCMC, which is equivalent to setting their frequencies to zero. 

\begin{figure}[t]
\centering
\includegraphics[width=0.4\textwidth]{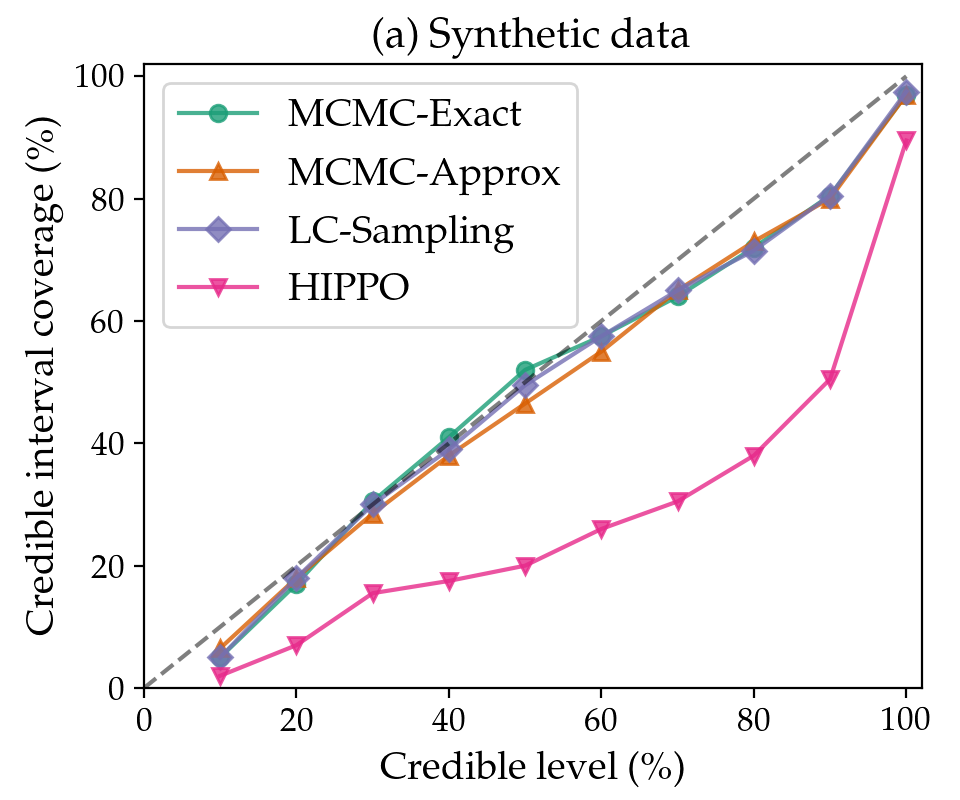}
\includegraphics[width=0.4\textwidth]{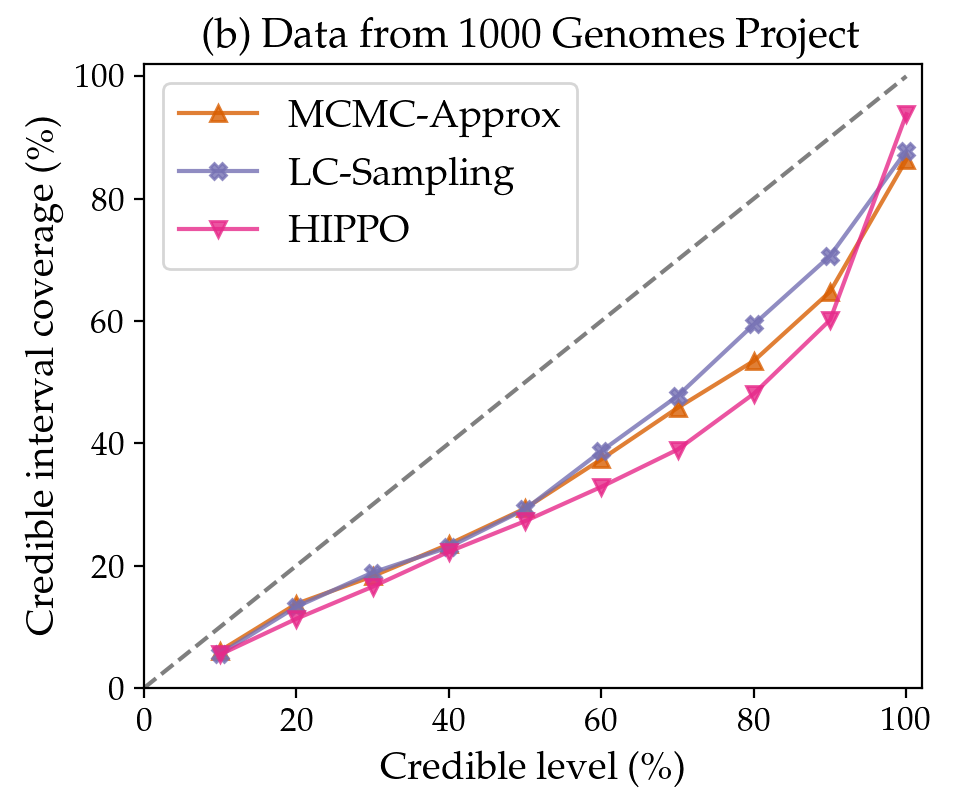}
\caption{\small Coverage of credible intervals for haplotype frequencies across (a)~25 synthetic datasets, (b)~100 datasets simulated based on 1KGP data. Input haplotypes that are absent from the population are excluded.} \label{fig:ci_coverage}
\end{figure}

To check if uncertainty is adequately accounted by the Bayesian methods, we report the coverage of (equal-tail) credible intervals of the haplotype frequencies for the synthetic datasets are shown in Figure~\ref{fig:ci_coverage}(a). The coverage of a $x\%$ credible interval is the proportion of haplotypes present in the population whose $x\%$ credible interval contains the corresponding true frequency. Our proposed methods give credible interval coverages that are close to the corresponding credible levels. The close agreement between MCMC-Exact and LC-Sampling is an indication that both methods produce the same posterior. The coverage for HIPPO is lower than expected, which is likely due to the removal of rare haplotypes during MCMC.

Out of the compared methods, AEML is the fastest, taking less than 1 second for each dataset. We report in Figure~\ref{fig:synthetic_shared_comp}(a) the runtimes (wall time) for the Bayesian methods, including any pre-processing steps (e.g. enumerating feasible sets for MCMC-Exact). The time taken by MCMC-Exact increases rapidly with pool size as the feasible sets get larger. There is considerable variation in runtime across datasets of the same pool size as the runtime is sensitive to the size of the feasible sets. The computational complexity of LC-Sampling is roughly linear with respect to pool size. The runtimes of MCMC-Approx and HIPPO are fairly insensitive to pool size, with MCMC-Approx being less than an order of magnitude slower than HIPPO.  In Figure~\ref{fig:synthetic_shared_comp}(b), we show boxplots of the ESS of haplotype frequencies, grouped by the pool size of each dataset. The ESS under MCMC-Exact and MCMC-Approximate are comparable, whereas the ESS under LC-Sampling decreases as pool size increases. Although HIPPO is the fastest Bayesian method, its ESS has the largest variation. In the worst case, its minimum ESS is close to the number of chains, indicating that chains are stuck in different modes of the posterior.

\subsection{Simulated haplotype data from 1000 Genomes Project}\label{sec:1k_genomes}

\begin{figure}[t]
\centering
\includegraphics[width=\textwidth]{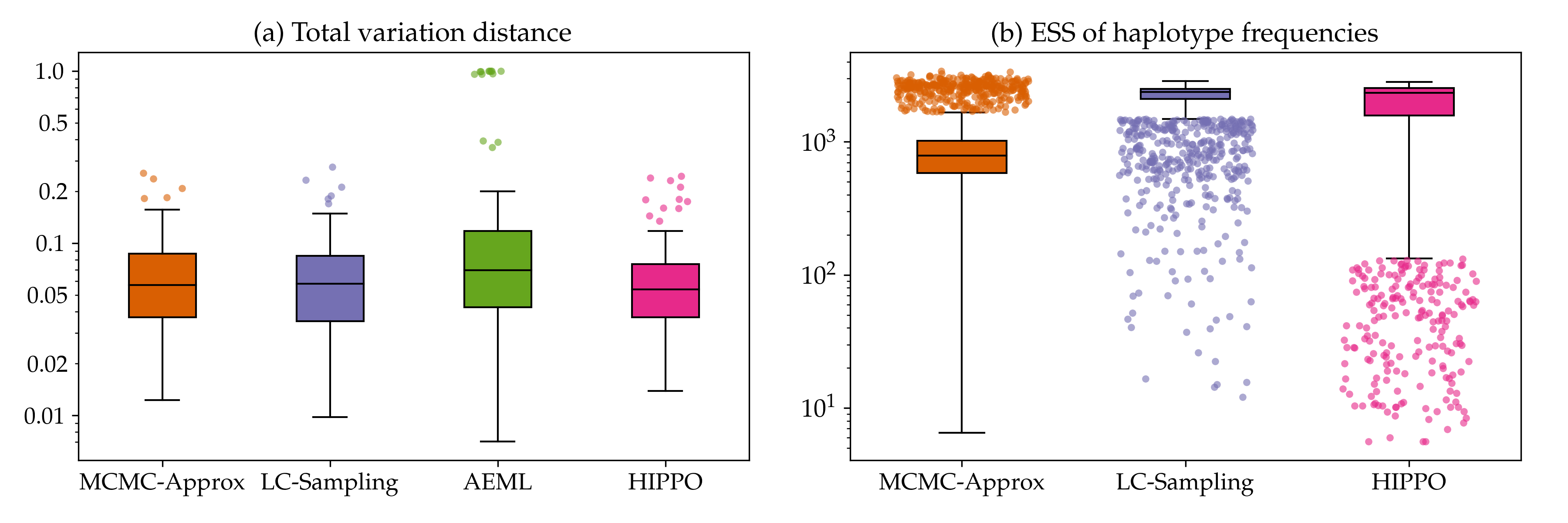}
\caption{\small (a)~TVD between point estimates (MLE for AEML, posterior mean for others) and the true haplotype frequencies across 100 datasets simulated based on data from 1KGP. (b)~ESS of haplotype frequencies across 100 datasets simulated based on data from 1KGP. Only haplotypes determined by partition ligation are included in the plot.} \label{fig:kgenomes_tvd_ess}
\end{figure}

We also compare our approach with existing methods based on data simulated with haplotype frequencies extracted from the 1000 Genomes Project (1KGP)~\parencite{the_1000_genomes_project_consortium_global_2015}. We use 190 unrelated haplotype samples of the CEU population (Utah residents with ancestry from Northern and Western Europe) for the region ENm010 on chromosome 7. This population and genetic region has been analysed by previous literature in haplotype inference for pooled genetic data \parencites{kirkpatrick_haplopool_2007,pirinen_estimating_2008,pirinen_estimating_2009,gasbarra_estimating_2011}. Following \textcite{gasbarra_estimating_2011}, we select the first 800 SNPs of the ENm010 region such that adjacent SNPs are separated by at least 100 base pairs. We construct 100 datasets by segmenting this sequence of 800 SNPs into $M=8$ SNPs (i.e. markers) per dataset. Each dataset consists of $N=20$ pools, each with $50$ haplotypes sampled with replacement from the 190 haplotype samples extracted from 1KGP. We exclude MCMC-Exact as the number of input haplotypes for some datasets is too large for feasible sets to be enumerated within reasonable time. 

The number of haplotypes present in each dataset ranges between 3 to 12, considerably smaller than $2^8=256$. We apply partition ligation (Section~\ref{sec:partition_ligation}) to obtain a list of input haplotypes for each dataset, which is used for all inference methods except for HIPPO, as HIPPO samples the list of input haplotypes as part of its MCMC procedure. For each dataset, the number of input haplotypes obtained from partition ligation ranges between 13 and 40. This implies that many input haplotypes have a true frequency of 0. We specify a sparser prior $\mathbf{p}\sim\mathrm{Dir}(0.1, \ldots, 0.1)$ for MCMC-Approx and LC-Sampling. For HIPPO, we keep the default Dirichlet concentration of $
\alpha=10^{-5}$, which is recommended~\parencite{pirinen_estimating_2009} as HIPPO implicitly considers all possible haplotypes. Out of the 100 lists produced by partition ligation, 43 of them included all haplotypes that are truly present. The sum of frequencies of haplotypes missed by partition ligation for each dataset averages to 0.0066, with the maximum frequency of such a haplotype being 0.0368. Since the number of input haplotypes for MCMC-Approx and LC-Sampling is not too large, we keep the same number of MCMC iterations for MCMC-Approx and LC-Sampling from Section~\ref{sec:sim_shared}. However, HIPPO implicitly considers all $256$ haplotypes, so we increase the number of MCMC iterations per chain from $5\times 10^5$ to $2.5\times 10^6$.

\begin{figure}[t]
\centering
\includegraphics[width=\textwidth]{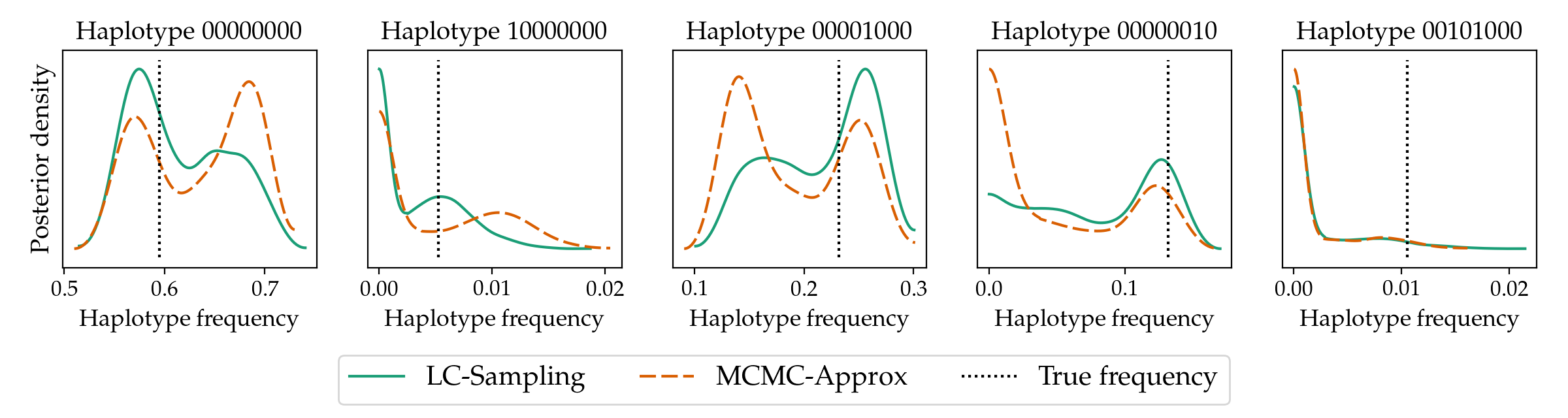}
\caption{\small Multimodal posterior distributions of selected haplotype frequencies from dataset 3  (based on 1KGP data). The posteriors under LC-Sampling and MCMC-Approx are shown as solid and dashed curves respectively; the true frequency is indicated by the vertical dotted line.} \label{fig:kgenomes_multimodal}
\end{figure}

The distributions of TVDs~\eqref{eq:tvd} across the 100 datasets between the true haplotype frequencies and point estimates under each method are shown in Figure~\ref{fig:kgenomes_tvd_ess}(a). AEML performs poorly on some datasets (TVD close to 1), possibly due to errors introduced by the normal approximation. The TVDs for the Bayesian methods are comparable, with LC-Sampling having a slightly lighter right tail. The average runtime for MCMC-Approx, LC-Sampling, AEML, and HIPPO are 2.2 minutes, 7.1 minutes, 0.2 minutes, and 6.3 minutes respectively. In Figure~\ref{fig:kgenomes_tvd_ess}(b), we show for each Bayesian method the ESS distribution of the frequencies of the haplotypes determined by partition ligation. Overall, chains from LC-Sampling exhibit the least autocorrelation. The Markov chains for MCMC-Approx and HIPPO become stuck at different modes for some haplotypes, as indicated by ESS values around 10. We find that for some datasets, there are multiple modes that are associated with comparable probability mass, see Figure~\ref{fig:kgenomes_multimodal} for a representative example. We note that the true frequency may or may not coincide with one of the modes. For this example, LC-Sampling and MCMC-Approx identify modes at similar frequencies, but the densities can be significantly different between methods. Posteriors under HIPPO are omitted as the inference model is different. Trace plots (Figure~\ref{fig:kgenomes_traces}) of these haplotype frequencies reveal that LC-Sampling and MCMC-Approx are able to switch efficiently between modes, whereas HIPPO tends to be stuck in one mode for a large number of iterations.

The coverage of credible intervals for all Bayesian methods are less than ideal (Figure~\ref{fig:ci_coverage}(b)), indicating that uncertainty is underestimated. For MCMC-Approx and LC-Sampling, the deterioration of coverage relative to Figure~\ref{fig:ci_coverage}(a) is attributed to the credible intervals not accounting for the uncertainty due to the input haplotype lists obtained via partition ligation. For example, the credible interval for a haplotype that is present in the population but missed by partition ligation is exactly zero, regardless of the credible level. Out of all Bayesian methods, the underestimation of uncertainty is least severe for LC-Sampling.

\subsection{Synthetic time-series data}\label{sec:tseries}

\begin{figure}[t!]
\centering
\small
\includegraphics[width=0.248\textwidth]{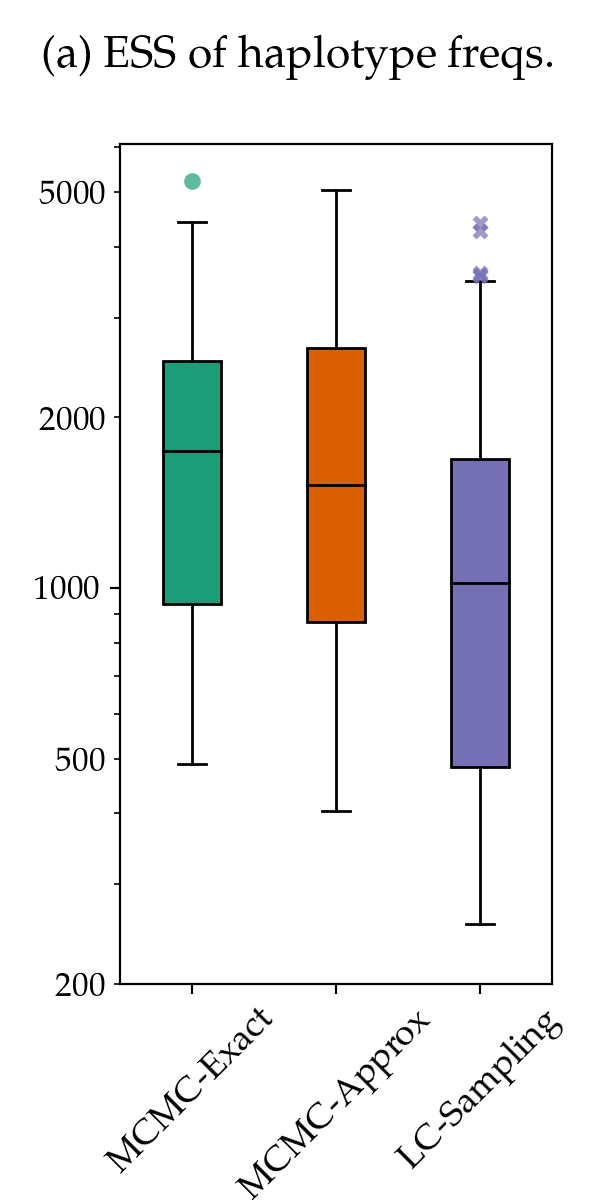}
\includegraphics[width=0.744\textwidth]{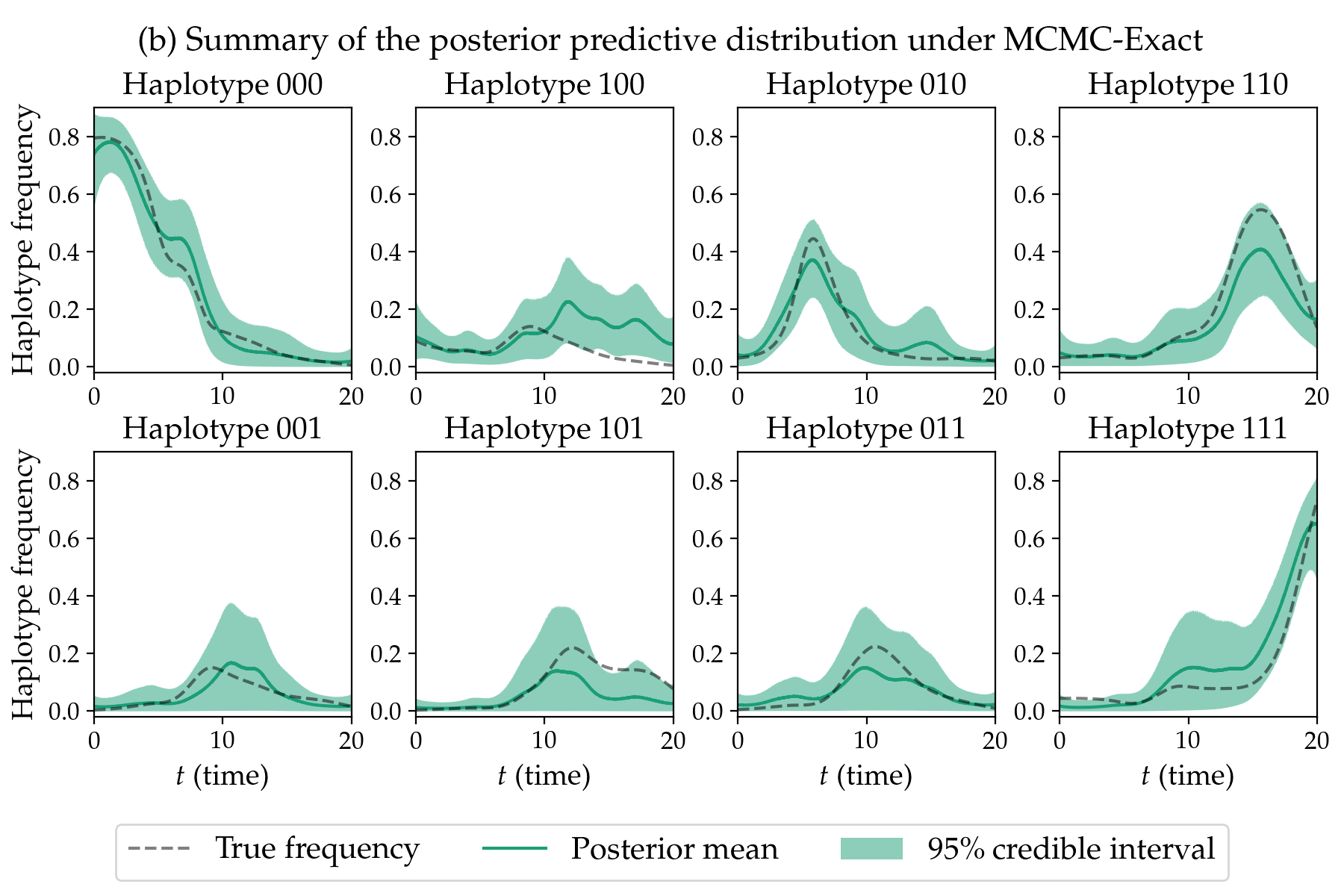}
\caption{\small (a)~Boxplots of ESS for haplotype frequencies $\{p_{ih}\}_{1\le i\le N, 1\le h\le H}$ under each proposed method for the time-series example. (b)~Posterior predictive distribution of haplotype frequencies under MCMC-Exact. The dashed and solid curves correspond to the true frequencies used for data simulation and the posterior mean respectively. Bands show 95\% credible intervals.} \label{fig:tseries_results}
\end{figure}

As an demonstration of how our methods extend to a hierarchical setting, we perform inference for a latent multinomial GP model applied to time-series data, as introduced in Section~\ref{sec:hier_ext}. To generate data, we simulate time-varying frequencies of $H=8$ haplotypes over $M=3$ markers from a differential equation system. We then simulate haplotype count data over $N=30$ time points with pool sizes of $n=50$ from a Dirichlet-multinomial distribution, and take the allele counts of each marker as the observed data. A Dirichlet-multinomial distribution is used to simulate overdispersion, whereas the inference model accounts for overdispersion through a white noise kernel (see~\eqref{eq:kernel_example}). The intention behind this mis-specification is to check whether our inference is robust against the overdispersion model. Details of the simulation and the complete inference model are given in Appendix~\ref{sec:tseries_apdix}.

We perform inference using our three proposed methods. Since the hierarchical model introduces correlations between model parameters, we increase the number of MCMC iterations performed (Table~S1). LC-Sampling requires more iterations as there is strong dependence between $\mathbf{z}_i$ and $\mathbf{p}_i$. Figure~\ref{fig:tseries_results}(a) shows that despite running LC-Sampling for 20 times more inference iterations, its MCMC output has lower ESS than MCMC-Exact and MCMC-Approx. Nevertheless, we did not encounter any MCMC convergence issues for the time-series data.

In Figure~\ref{fig:tseries_results}(b), we plot the posterior predictive distribution under MCMC-Exact. There is general agreement between the posterior means and the true haplotype frequencies, with the caveat that the posterior accounts for noise, but the true frequencies are not perturbed by noise. We note that the credible intervals for the haplotypes in the bottom row of Figure~\ref{fig:tseries_results}(b) have wide credible intervals around $t=10$. Closer inspection reveals that this is caused by posterior multimodality and parameter non-identifiability due to insufficient signal in the data (Appendix~\ref{sec:tseries_apdix}). We report the posterior predictive distributions under MCMC-Approx and LC-Sampling in Figures~\ref{fig:tseries_mn_trends} and \ref{fig:tseries_gibbs_trends}, which are highly similar to that of MCMC-Exact.

\section{Discussion} \label{discussion}

In this paper, we have developed two exact methods (MCMC-Exact and LC-Sampling) and an approximate method (MCMC-Approx) for Bayesian inference of haplotype frequencies given pooled genotype data under a latent multinomial model. The latent multinomial framework is suitable for handling incomplete reporting of genetic data, as full haplotype information is not always available. Furthermore, we illustrate how our methods can infer haplotype frequencies of multiple related populations with a hierarchical model. Existing statistical methods either have only been applied to small pool sizes ($n \le 20$), or rely on approximations. However, approximate methods may give unreliable inference when applied to real data. We instead recommend the use of MCMC-Exact for problems that are small enough where enumerating feasible sets is practical, and LC-Sampling for larger problems.

Out of our proposed methods, MCMC-Approx is the fastest as its runtime is relatively insensitive to pool size (Figure~\ref{fig:synthetic_shared_comp}). However, its performance is less consistent than the exact methods --- we find good agreement between the results from MCMC-Approx and LC-Sampling only for our synthetic data examples (Section~\ref{sec:sim_shared} and \ref{sec:tseries}). For datasets simulated from real genetic data (Section~\ref{sec:1k_genomes}), there are 8 markers per dataset, but only 3 to 12 haplotypes that are truly present in each dataset. Thus, some datasets have markers with highly correlated allele counts, resulting in near-singular covariance matrices where the approximate likelihood has a larger curvature than the exact likelihood (see Figure~\ref{fig:mn_vs_exact_mle}). This explains why for MCMC-Approx, the Markov chains do not converge in some cases, and uncertainty is more severely underestimated compared to the exact method LC-Sampling. This is also a likely reason for why AEML, a maximum likelihood method, fails on some of these datasets. We speculate that normal approximation methods may be reliable if one is confident that all haplotype frequencies are nonzero, but further investigation into this is needed.

Turning to exact methods, we find that the enumeration method MCMC-Exact does not scale well with pool size, as the size of the feasible set grows rapidly. LC-Sampling addresses this issue by sampling Markov chains over the feasible set, without resorting to approximations. The computational savings come from exploring only a subset of the feasible set that is likely to produce the observed data. The parameter $C$ (Algorithm~\ref{algo:mwg_dirmult}) gives us control over how the runtime of LC-Sampling scales. However, LC-Sampling produces Markov chains that exhibit more autocorrelation, especially as pool size increases (Figure~\ref{fig:synthetic_shared_comp}(b)). The reason for this is twofold: the number of latent count values for MCMC to explore becomes greater, and the conditional posterior~\eqref{eq:dir_post} from which the frequencies are sampled becomes more influenced by the likelihood than the prior. The posterior samples of the frequencies become more dependent on the latent counts, thereby increasing autocorrelation. For the time-series example, LC-Sampling also gave the lowest ESS, as the alternating updates of strongly dependent variables $\mathbf{z}_i$ and $\mathbf{p}_i$ ($i=1,\ldots,N$) give rise to greater autocorrelation~\parencite{hills_parameterization_1992}.

Interestingly, MCMC-Approx gives lower ESS than LC-Sampling for datasets simulated based on real genetic data. One explanation is that MCMC-Approx overestimates the density at some posterior modes (see bottom right plot of Figure~\ref{fig:mn_vs_exact_mle}), which makes it more difficult for a chain to switch between modes. Multimodal posteriors are notoriously difficult for MCMC methods to sample. When faced with a multimodal posterior, a single chain produced by HIPPO may not switch between modes even after millions of iterations (Figure~\ref{fig:kgenomes_traces}). To address this, \textcite{pirinen_estimating_2009} proposed to only keep the chain whose posterior mean has the highest posterior density. This is sensible if most of the posterior mass is concentrated around one sharp mode. Unfortunately, this is not the case, as multimodal posteriors often have modes with comparable posterior mass, e.g. Figure~\ref{fig:kgenomes_multimodal} and Figures~\ref{fig:tseries_exact_joint}--\ref{fig:tseries_gibbs_joint}. Keeping only one chain that is stuck at the global mode does not properly account for uncertainty. Moreover, it is possible that the true frequencies may not even occur near the global mode. We also note that maximum likelihood methods that optimise towards a single mode, such as AEML, would fail to account for uncertainty across multiple modes. In contrast, exact Bayesian methods are able to produce inference that is robust against multimodality.

In comparison to HIPPO, our proposed methods give more reliable estimates of uncertainty (Figure~\ref{fig:ci_coverage}), and give smaller errors in the case where all input haplotypes are known (Figure~\ref{fig:synthetic_shared_errors}). However, our proposed methods may miss some haplotypes if the input list is determined via partition ligation, which occurred for 57 out of the 100 1KGP datasets. Nevertheless, our posterior means still achieve TVDs that are no worse than HIPPO. A potential alternative is to replace the MCMC-Approx subroutine in partition ligation with sparse optimisation methods for frequency estimation \parencite{jajamovich_maximum-parsimony_2013,zhou_cshap_2019}.

Other inference methods for latent multinomial models have been proposed in literature outside from haplotype inference. An alternative to the Markov basis we use in LC-Sampling is the dynamic Markov basis~\parencite{bonner_dmb,hazelton_dmb}, which determines proposal directions on-the-fly during MCMC. For large configuration matrices, a Markov basis may be too large to be practically computed, whereas a dynamic Markov basis uses a relatively small number of proposal directions that depend on the current value of the latent counts during MCMC. The method guarantees that the resulting Markov chain over latent counts is irreducible, but requires expert implementation \parencite{zhang_fast_2019}. We are also aware of the saddlepoint approximation as an alternative to the normal approximation for the latent multinomial model \parencite{zhang_fast_2019}. However, we suspect that this approximation suffers from similar issues as MCMC-Approx, as it uses a Hessian matrix that shares similar structure with the covariance matrix used in the normal approximation~\eqref{eq:mn_approx}.

Compared to existing approaches, the methods that we propose in this paper for haplotype inference from pooled genetic data are more widely applicable. The implementation of the existing methods AEML and HIPPO assume that the data consists of allele counts of each genetic marker. Our methods only require each count to correspond to a subset of the full haplotypes, and these subsets can vary across pools. For example, a study may report complete haplotype information on a subset of the genetic markers. Moreover, we have implemented our methods using the probabilistic programming library \texttt{PyMC}~\parencite{Salvatier2016}, such that the methods can be easily extended to hierarchical settings, as demonstrated in \ref{sec:tseries}. In future work, we will apply our methods to spatiotemporal modelling of antimalarial drug resistance. In particular, we are interested in resistance against the antimalarial sulfadoxine-pyrimethamine (SP) for the parasite \textit{Plasmodium falciparum}. This resistance is characterised by specific mutations on the \textit{dhfr} and \textit{dhps} genes~\parencite{sibley_sp}, and reporting inconsistencies between genetic studies has been previously noted~\parencite{ebel_angola}. Our methods developed in this paper applied to a hierarchical model can readily handle such inconsistencies to produce predictive spatiotemporal maps for the prevalences of SP-resistant haplotypes.

\section{Acknowledgements}
 J.A. Flegg’s research is supported by the Australian Research Council (DP200100747, FT210100034) and the National Health and Medical Research Council (APP2019093). 

\emergencystretch 2em
\printbibliography[heading=bibintoc] 

\pagebreak
\begin{appendices}

\setcounter{figure}{0}
\setcounter{table}{0}

\renewcommand{\thetable}{A\arabic{table}}%
\renewcommand{\thefigure}{A\arabic{figure}}%

\makeatletter
\renewcommand{\thealgocf}{A\@arabic\c@algocf}
\renewcommand{\fnum@algocf}{\AlCapSty{\AlCapFnt\algorithmcfname\nobreakspace\thealgocf}}%
\makeatother

\begin{refsection}

\section{Algorithm for finding the feasible set}\label{sec:bnb}

In this section, we describe a branch-and-bound algorithm for solving $\mathbf{A}\mathbf{z}=\mathbf{y}$ over nonnegative integers $z_1,\ldots,z_H$ given a binary matrix $\mathbf{A}\in\{0,1\}^{R\times H}$ and nonnegative integers $y_1,\ldots,y_R$. Note that the index $i$ from the main text is dropped for conciseness here. We assume that the condition $z_1+\cdots+z_H=n$ is encoded in the linear system $\mathbf{A}\mathbf{z}=\mathbf{y}$, and that the configuration matrix $\mathbf{A}$ is of full row rank (see Section~2.2 of main text). Since $\mathbf{A}$ is of full row rank, we can find $R$ columns of $\mathbf{A}$ that are linearly independent. Without loss of generality, we rearrange the columns of $\mathbf{A}$ such that these $R$ linearly independent columns are the last $R$ columns, denoted as $\mathbf{A}_{H-R+1:H}$. Since $\mathbf{y} = \mathbf{A}_{1:H-R}\mathbf{z}_{1:H-R} + \mathbf{A}_{H-R+1:H}\mathbf{z}_{H-R+1:H}$, it follows that
\begin{equation}\label{eq:zsuff_inv}
     \mathbf{z}_{H-R+1:H} = \mathbf{A}_{H-R+1:H}^{-1}(\mathbf{y}-\mathbf{A}_{1:H-R}\mathbf{z}_{1:H-R}),
\end{equation}
where $\mathbf{A}_{1:H-R},\mathbf{z}_{1:H-R},\mathbf{z}_{H-R+1:H}$ denotes the first $H-R$ columns of $\mathbf{A}$, the first $H-R$ entries of $\mathbf{z}$, the last $R$ entries of $\mathbf{z}$ respectively. To find all solutions to the system, we perform a branch-and-bound search to find all possible values of $z_1,\ldots, z_{H-R}$. Starting from $h=1$, the algorithm branches on an interval of possible values for $z_h$ and increments $h$ whenever a branch is travelled down. If this succeeds until $h=H-R$, we then find the last $R$ entries of $\mathbf{z}$ by using \eqref{eq:zsuff_inv}. If the result consists of nonnegative integers, we accept $\mathbf{z}$ as a solution to $\mathbf{A}\mathbf{z}=\mathbf{y}$. We then backtrack the search path (decrementing $h$), and explore all other branches to find other solutions. The search is made efficient by finding lower and upper bounds for $z_h$ based on the values of $z_1,\ldots,z_{h-1}$ when branching on the value of $z_h$ for $h=1,\ldots,H$.

\setcounter{algocf}{0}
\begin{algorithm}[t]
\DontPrintSemicolon
\small
\KwIn{$\mathbf{y},\mathbf{A},n,l_1,\ldots,l_H,u_1,\ldots,u_H$}
\KwOut{$\mathcal{S}$, a set of nonnegative integer solutions $\mathbf{z}$ to $\mathbf{A}\mathbf{z}=\mathbf{y}$}
$\mathcal{S} \gets \{\}$\;
$\mathbf{z} \gets \text{empty vector of size $H$}$\;
$U_1,U_0,L_1,L_0 \gets \text{empty $R\times(H-R)$ array}$\;
\For{$r\gets 1$ \KwTo $R$}{
    $U_1[r,1] \gets y_r-\sum_{h=2}^H a_{r,h}l_h$\; 
    $U_0[r,1] \gets n-y_r-\sum_{h=2}^H (1-a_{r,h})l_h$\;
    $L_1[r,1] \gets y_r-\sum_{h=2}^H a_{r,h}u_h$\;
    $L_0[r,1] \gets n-y_r-\sum_{h=2}^H (1-a_{r,h})u_h$\;
}
compute $\mathbf{A}_{H-R+1:H}^{-1}$\;
\SetKwFunction{FnBranch}{branch}
\SetKwProg{Fn}{Function}{:}{}
\Fn{\FnBranch{$h$}}{
    $z_\text{min} = \max\!\left(l_h, \max_{r=1,\ldots,R} L_{a_{r,h}}[r,h]\right)$\;
    $z_\text{max} = \min\!\left(u_h, \min_{r=1,\ldots,R} U_{a_{r,h}}[r,h]\right)$\;
    \If{$h=H-R$}{
        \For{$z_h\gets z_\text{min}$ \KwTo $z_\text{max}$}{
            $\mathbf{z}_{H-R+1:H} \gets \mathbf{A}_{H-R+1:H}^{-1}(\mathbf{y}-\mathbf{A}_{1:H-R}\mathbf{z}_{1:H-R})$\;
            \If{$\text{all entries of } \mathbf{z}_{H-R+1:H} \text{ are nonnegative integers}$}{
                $\mathcal{S}\gets \mathcal{S}\cup \{\mathbf{z}\}$\;
            }
        }
    }
    \Else{
        \For{$z_h\gets z_\text{min}$ \KwTo $z_\text{max}$}{
            \For{$r\gets 1$ \KwTo $R$}{
                $U_1[r,h+1] \gets U_1[r,h]-a_{r,h}z_h+a_{r,h+1}l_{h+1}$\;
                $U_0[r,h+1] \gets U_0[r,h]-(1-a_{r,h})z_h+(1-a_{r,h+1})l_{h+1}$\;
                $L_1[r,h+1] \gets L_1[r,h]-a_{r,h}z_h+a_{r,h+1}u_{h+1}$\;
                $L_0[r,h+1] \gets L_0[r,h]-(1-a_{r,h})z_h+(1-a_{r,h+1})u_{h+1}$\;
            }
            \FnBranch{$h+1$}\;
        }
    }
    \KwRet\;
}
\FnBranch{$1$}\;
\KwRet{$\mathcal{S}$}\;
\caption{\small Branch-and-bound search for integer linear system with 0-1 coefficients over nonnegative integers with known sum}\label{algo:zsolve}
\end{algorithm}

Before the search procedure, we first determine preliminary lower bounds $l_h$ and upper bounds $u_h$ for each entry $z_h$  that are satisfied by all nonnegative integer solutions to $\mathbf{A}\mathbf{z}=\mathbf{y}$. A simple choice is to set
\begin{equation}
    l_h = 0, \qquad
    u_h = \min\limits_{r=1,\ldots,R} \{a_{r,h}y_r + (1-a_{r,h})(n-y_r)\}.
    \label{eq:pre_bounds}
\end{equation}
The lower bound is trivial, whereas the upper bound is true because the $r$-th equation in the system implies that $z_h \le y_r$ if $a_{r,h}=1$, or $z_h \le n-y_r$ if $a_{r,h}=0$. For each $h=1,\ldots,H$, we now seek to derive bounds for $z_h$ using  the values of $z_1,\ldots,z_{h-1}$ along the current search path. For any fixed $r$ and $h$, we have
\begin{align}
    z_h &= l_h + (z_h-l_h) \nonumber\\
        &\le l_h + \sum_{h'=h}^H \mathds{1}(a_{r,h'}=a_{r,h})(z_{h'}-l_{h'}) \nonumber\\
        &=
        \begin{cases}
        \displaystyle y_r - \sum_{h'=1}^{h-1} a_{r,h'}z_{h'} - \smashoperator{\sum_{h'=h+1}^H} a_{r,h'}l_{h'} &\text{if }a_{r,h}=1,\\
        \displaystyle n - y_r - \sum_{h'=1}^{h-1} (1-a_{r,h'})z_{h'} - \smashoperator{\sum_{h'=h+1}^H} (1-a_{r,h'})l_{h'} &\text{if }a_{r,h}=0.\\
        \end{cases} \label{eq:updated_ub}
\end{align}
The inequality holds since $z_{h'}\ge l_{h'}$, while the last equality holds because of $y_r = \sum_{h'=1}^H a_{r,h'}z_{h'}$ and $n-y_r = \sum_{h'=1}^H (1-a_{r,h'})z_{h'}$. We define
\begin{align*}
    U_1(r; h, z_1,\ldots z_{h-1}) &= y_r - \sum_{h'=1}^{h-1} a_{r,h'}z_{h'} - \smashoperator{\sum_{h'=h+1}^H} a_{r,h'}l_{h'}, \\
    U_0(r; h, z_1,\ldots z_{h-1}) &= n - y_r - \sum_{h'=1}^{h-1} (1-a_{r,h'})z_{h'} - \smashoperator{\sum_{h'=h+1}^H} (1-a_{r,h'})l_{h'},
\end{align*}
to write the inequality in \eqref{eq:updated_ub} more concisely as $z_h \le U_{a_{r,h}}(r; h, z_1,\ldots z_{h-1})$. We similarly define
\begin{align*}
    L_1(r; h, z_1,\ldots z_{h-1}) &= y_r - \sum_{h'=1}^{h-1} a_{r,h'}z_{h'} - \smashoperator{\sum_{h'=h+1}^H} a_{r,h'}u_{h'}, \\
    L_0(r; h, z_1,\ldots z_{h-1}) &= n - y_r - \sum_{h'=1}^{h-1} (1-a_{r,h'})z_{h'} - \smashoperator{\sum_{h'=h+1}^H} (1-a_{r,h'})u_{h'},
\end{align*}
to obtain the inequality $z_h \ge L_{a_{r,h}}(r; h, z_1,\ldots z_{h-1})$. 

The branch-and-bound algorithm is given in Algorithm~\ref{algo:zsolve}. The values for $U_1,U_0,L_1,L_0$ are initialised in lines 4--8, where $h=1$ and $r=1,\ldots,R$. Given the values of $z_1,\ldots,z_{h-1}$ on the current search path, the algorithms finds lower and upper bounds for $z_h$ in lines 11--12 using the inequality $L_{a_{r,h}}(r; h, z_1,\ldots z_{h-1}) \le z_h \le U_{a_{r,h}}(r; h, z_1,\ldots z_{h-1})$ over $r=1,\ldots,R$. The branching occurs in lines 19--24, where $U_1,U_0,L_1,L_0$ are updated based on the chosen value of $z_h$.

If the actual range of values that $z_h$ can take is much narrower than the interval $[l_h,u_h]$ as defined in \eqref{eq:pre_bounds}, it may be computationally more efficient to find the actual minimum and maximum values that $z_h$ can take, i.e. setting
\begin{equation}
\begin{split}
    l_h &= \min \{z_h : \mathbf{A}\mathbf{z}=\mathbf{y}, z_1\ge 0, \ldots, z_H\ge 0\}, \\
    u_h &= \max \{z_h : \mathbf{A}\mathbf{z}=\mathbf{y}, z_1\ge 0, \ldots, z_H\ge 0\}.
\end{split}
\label{eq:optim}
\end{equation}
for each $h=1,\ldots,H$. These optimisation problems can be solved using integer linear programming. This introduces a computational overhead before the branch-and-bound search, but prunes the search space as $\mathbf{z}$ would have tighter bounds.

\section{Multimodality example from 1000 Genomes Project}

\begin{figure}[t]
\centering
\includegraphics[width=\textwidth]{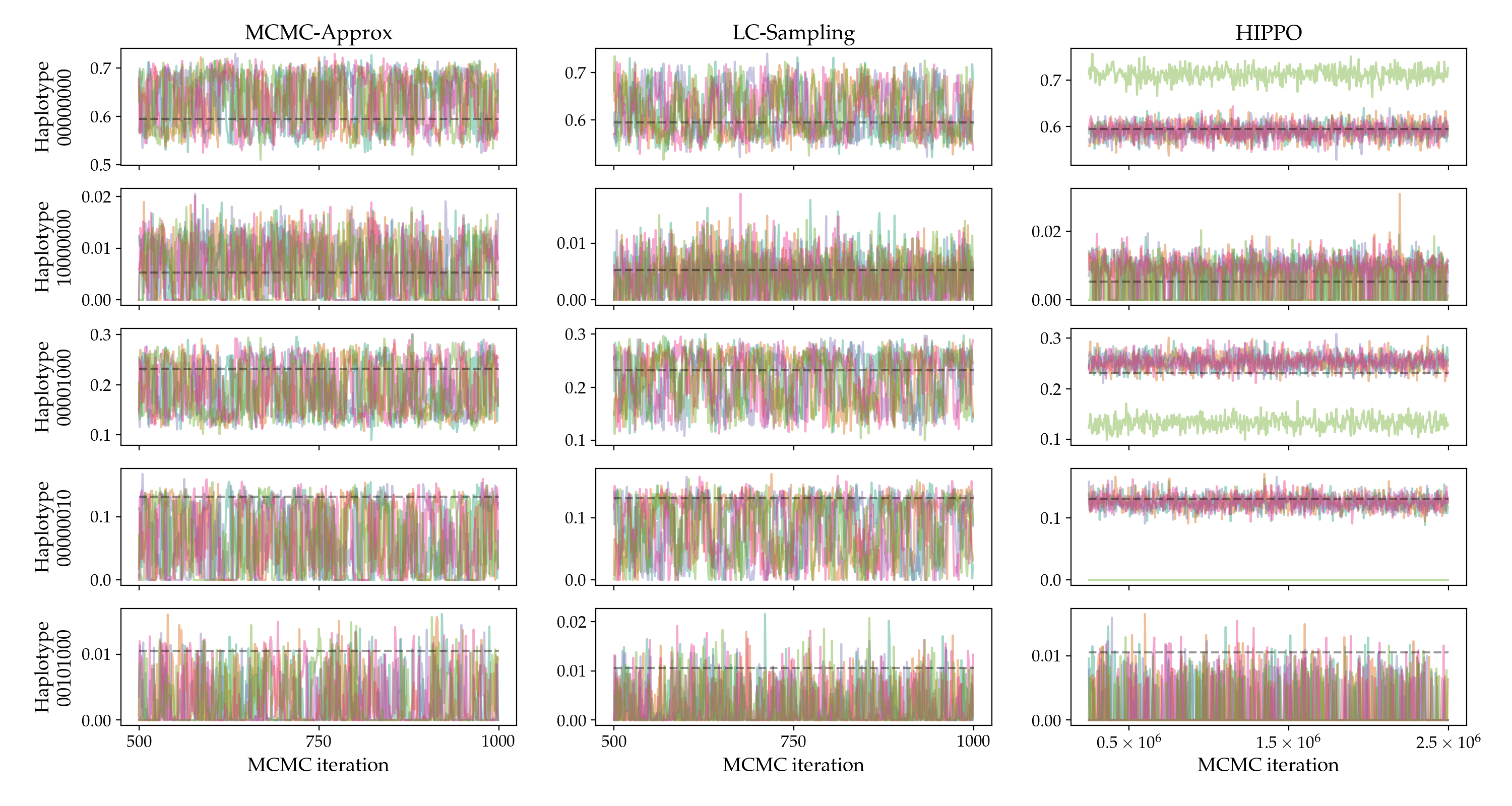}
\caption{\small Trace plots of selected haplotype frequencies that depict posterior multimodality from dataset 3 simulated based on genetic data from the 1000 Genomes Project, with burn-in iterations excluded and thinning applied for HIPPO. The dashed line corresponds to the true haplotype frequency.} \label{fig:kgenomes_traces}
\end{figure}

In Figure~7 of the main text, we give an example of posterior multimodality when fitting the latent multinomial model to a dataset simulated based on genetic data from the 1000 Genomes Project. The trace plots of the corresponding haplotype frequencies are given in Figure~\ref{fig:kgenomes_traces}. Note that a thinning factor of 4500 is applied for HIPPO. For this example, LC-Sampling exhibits the best MCMC mixing, followed by MCMC-Approx. HIPPO produces Markov chains that are stuck at different local modes for a long duration. In row 4, one of the chains neglects a haplotype with true frequency 0.13. In rows 2 and 5, the the support of each chain consists of a short interval close to zero and a longer interval away from zero, yet the true value is barely covered by the longer interval. The poor mixing of HIPPO chains may lead to inaccurate estimation. These conclusions drawn from our visual inspection of the trace plots are consistent with the lowest ESS of the
haplotype frequencies under each method: 371 for MCMC-Approx, 515 for LC-Sampling, and 10 for HIPPO.

\section{Additional details for time-series modelling}\label{sec:tseries_apdix}

\begin{figure}[t]
\centering
\includegraphics[width=\textwidth]{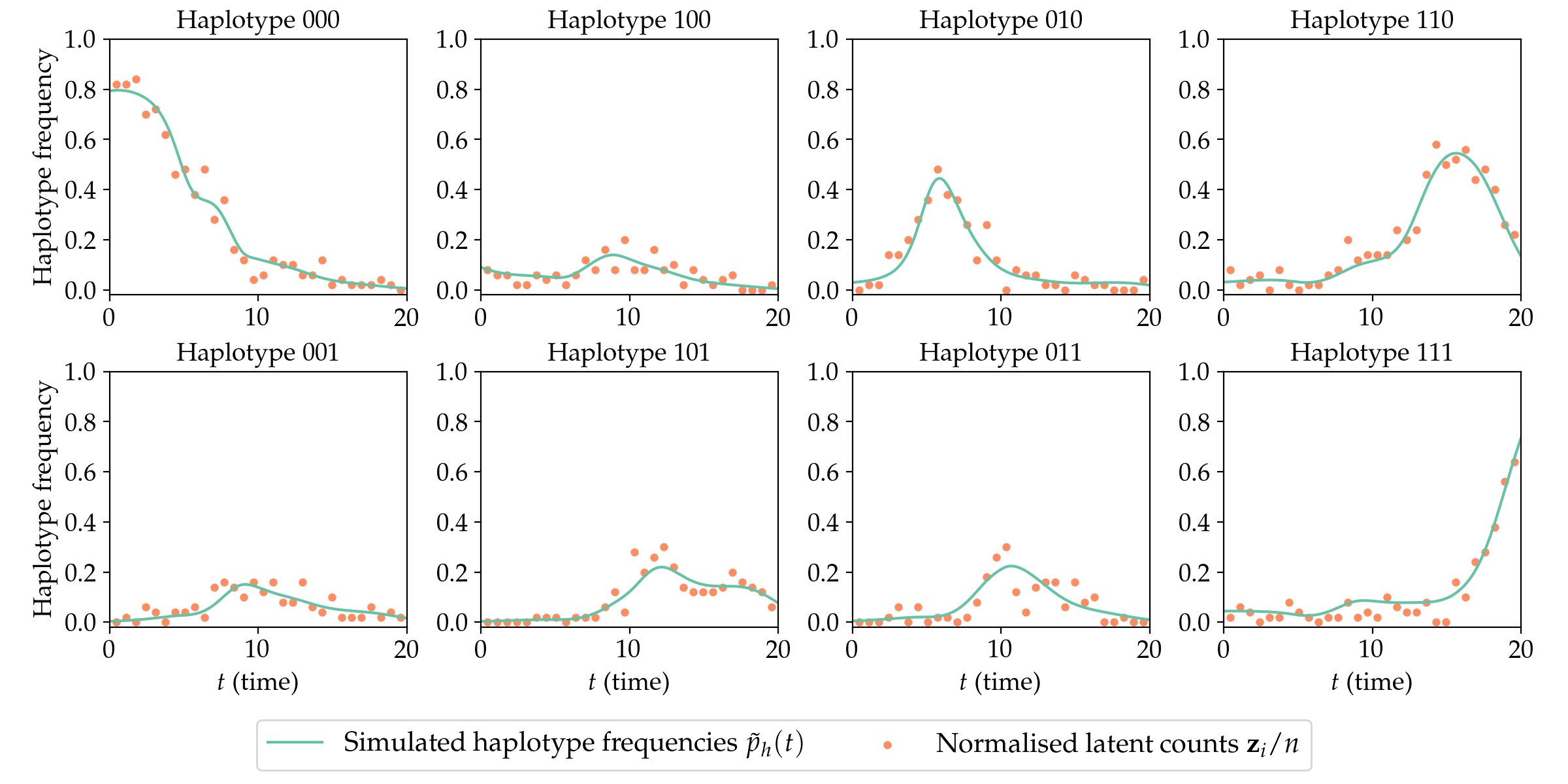}
\caption{\small True haplotype frequencies $\tilde{p}_1(t),\ldots,\tilde{p}_H(t)$ (solid curve) used for simulating latent counts (normalised, scatter points) for synthetic time-series data.} \label{fig:tseries_hap}
\end{figure}

We use a custom system of differential equations to simulate time-series of haplotype frequencies. The system is analogous to the continuous-time model of haploid selection expounded by \citet{hartl_popgen}, but extended for multiple haplotypes instead of two genotypes. Consider a population of malaria parasites each with one of $H=8$ possible haplotypes over 3 markers. For each $h=1,\ldots,H$, the number of parasites with haplotype $h$ at time $t$ is $N_h(t)$. We define the frequency of haplotype $h$ to be $\tilde{p}_h(t) \coloneqq N_h(t) / \sum_{h'=1}^H N_{h'}(t)$. Assuming exponential growth, we have $N'_h(t) = r_h(t)N_h(t)$, where $r_h(t)$ is a time-varying intrinsic growth rate for haplotype $h$, which we interpret as a measure of relative fitness, e.g. a drug-resistant haplotype has a higher fitness relative to a drug-sensitive haplotype after exposure to the drug. We set each $r_h(t)$ to be a sum of $D=4$ sigmoid functions:
\begin{equation}
    r_h(t) = \alpha_{h,0} + \sum_{d=1}^D \frac{\alpha_{h,d}-\alpha_{h,d-1}}{1+\exp(-(t-c_{h,d})/\gamma_{h,d})}.
\end{equation}
The $d$-th sigmoid ($d=1,\ldots,D$) for haplotype $h$ suggests some change in its relative fitness due to epidemiology or drug usage, which the changepoint occuring at $t=c_{h,d}$. We also impose the constraint $c_{h,1}<\cdots<c_{h,D}$. The coefficient $\gamma_{h,d}$ ($d=1,\ldots,D$) controls how quickly the change at $c_{h,d}$ occurs, whereas the coefficient $\alpha_{h,d}$ ($d=0,\ldots,D$) is the steady-state relative fitness between changepoints $c_{h,d}$ and $c_{h,d+1}$, where we define $c_{h,0}=0$ as the start point and $c_{h,D}=20$ as the end point. The coefficients of the sigmoid functions are sampled as follows:
\begin{align}
    c_{h,d} &\sim \mathrm{Uniform}(0, 20) & \text{for } h=1,\ldots,H,\, d&=1,2,3,4 \\
    \gamma_{h,d} &\sim \mathrm{Uniform}(0.2, 2.0) & \text{for } h=1,\ldots,H,\, d&=1,2,3,4 \\
    \alpha_{h,d} &\sim \mathrm{N}(0, 1/(c_{h,d+1}-c_{h,d})^2) & \text{for } h=1,\ldots,H,\, d&=0,1,2,3,4  \label{eq:alpha_sim}
\end{align}
For each fixed $h=1, \ldots, H$, we reorder $\{c_{h,d}\}_{d=1}^D$ such that the sequence is in increasing order. The normal standard deviation in \eqref{eq:alpha_sim} is inversely proportional to the distance between changepoints to discourage dramatic growth in $N_h(t)$ between two changepoints that are far apart. We choose the starting values $N_h(0)$ such that the median values of $N_h(t)$ over $t\in[0,20]$ are equal across $h=1,\ldots,H$.

\begin{figure}[t]
\centering
\includegraphics[width=\textwidth]{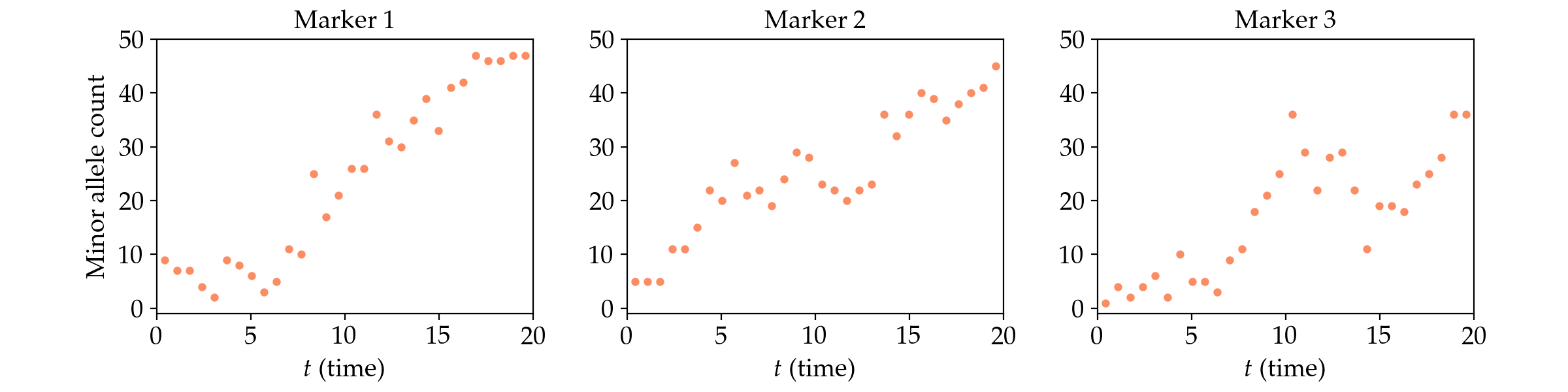}
\caption{\small Synthetic time-series data in the form of allele counts for 3 markers.} \label{fig:tseries_marker}
\end{figure}

We find that the resulting trends of $\tilde{p}_h(t)$ following the simulation above may be uninteresting depending on the random generation. For example, a haplotype may completely dominate the population, or too many haplotypes exhibit very little variation over time. To counter this, we carry out 100 simulations where  $\lvert 
\frac{d}{dt} \tilde{p}_{h}(t)\rvert < 1$ for all $t\in [0,20]$ (avoid domination), and select the simulation with the most temporal variation for generating the synthetic time-series count data. We quantify temporal variation using the heuristic
\begin{equation}
    \sum_{h=1}^H \sum_{t'=5}^{14} \lvert \tilde{p}_h(t'+1)-\tilde{p}_h(t') \rvert.
\end{equation}
The selected simulation is shown in Figure~\ref{fig:tseries_hap}, along with the $N=30$ latent counts $\{\mathbf{z}_i\}_{i=1}^N$ divided by the pool size $n=50$. The latent counts are overdispersed counts following the Dirichlet-multinomial distribution
\begin{align}
\mathbf{z}_i &\sim \mathrm{DirMult}(50, (200p_1(t_i), \ldots, 200p_H(t_i))), &\text{for }i=1,\ldots,N,
\end{align}
where $t_i = 0.66i - 0.23$ ($i=1,\ldots,30$) are equally spaced time points. The Dirichlet-multinomial distribution chosen has the same mean as $\mathrm{Mult}(50, (p_1(t_i), \ldots, p_H(t_i)))$, but with $24\%$ larger variance. Finally, the observed data are the allele counts of each marker across the time points $t_1,\ldots,t_N$, which is shown in Figure~\ref{fig:tseries_marker}.

For the latent multinomial GP model, we first define the haplotype frequencies $\mathbf{p}_1,\ldots,\mathbf{p}_N$ as a softmax transformation of Gaussian processes $f_1,\ldots,f_H$ observed at time points $\mathbf{t}\coloneqq (t_1,\ldots,t_N)$:
\begin{align}
p_{ih} &= \frac{\exp(f_h(t_i))}{\exp(f_1(t_i))+\cdots+\exp(f_H(t_i))} & \text{for } i=1,\ldots,N, \, h=1,\ldots,H.
\end{align}
Following Section~2.4 of the main text, we choose the mean function of the $h$-th GP to be a constant $\mu_h$, and the covariance function of the $h$-th GP to be the sum of a rational quadratic kernel and a white noise kernel,
\begin{equation}
c_h(t_i,t_{i'}) = s_h^2 \! \left( 1 + \frac{(t_i - t_{i'})^2}{2\tau_h^2} \right)^{-1} \hspace{-0.5em} + \sigma^2 \mathds{1}(i=i'),
\end{equation}
where $c_h(t_i,t_{i'})$ is the $(i,i')$-th entry of the covariance matrix $\mathbf{C}_h(\mathbf{t},\mathbf{t})$ for $f_h(\mathbf{t}) \coloneqq (f_h(t_1), \ldots, f_h(t_N))$, $\tau_h$ is the timescale, $s_h$ is the temporal standard deviation, $\sigma$ is the noise standard deviation, and $\mathds{1}(\cdot)$ is the indicator function. The full inference model is as follows:
\begin{align}
    \mathbf{y}_i &= \mathbf{A}_i\mathbf{z}_i & \text{for }i=1,\ldots N, \\
    \mathbf{z}_i \mid \mathbf{p}_i &\sim \mathrm{Mult}(n_i, \mathbf{p}_i) & \text{for } i=1,\ldots,N, \\
    f_h(\mathbf{t}) \mid \mu_h,s_h,\tau_h,\sigma &\sim \mathrm{N}(\mu_h \mathbf{1}_N, \mathbf{C}_h(\mathbf{t},\mathbf{t})) & i,i'=1,\ldots,N, \, h=1,\ldots,H, \\
    \bm\mu &\sim \mathrm{N}\!\left(\mathbf{0}_H, 2^2\left(\mathbf{I}_H-\frac{1}{H}\mathbf{J}_H\right)\right), \label{eq:sum_to_zero_normal} \\
    s_h &\sim \mathrm{InverseGamma}(3, 3) & \text{for } h=1,\ldots,H, \label{eq:s_prior} \\
    \tau_h &\sim \mathrm{InverseGamma}(3, 5) & \text{for } h=1,\ldots,H, \\
    \sigma &\sim \mathrm{InverseGamma}(3, 1), \label{eq:sigma_prior}
\end{align}
where $\bm\mu=(\mu_1,\ldots,\mu_h)$, $\mathbf{1}_N$ is a vector of $N$ ones, $\mathbf{0}_H$ is a vector of $H$ zeros, $\mathbf{I}_H$ is the $H\times H$ identity matrix, and $\mathbf{J}_H$ is a $H\times H$ matrix of ones. Note that if all entries of $\mu_h$ across $h=1,\ldots,H$ are incremented by the same value, this keeps the values of $\mathbf{p}_1,\ldots,\mathbf{p}_N$ unchanged. To improve identifiability of $\bm\mu$, we impose a sum-to-zero constraint $\mu_1+\cdots+\mu_H=0$ through the covariance matrix in \eqref{eq:sum_to_zero_normal}. For the nonnegative hyperparameters, we choose inverse gamma priors \eqref{eq:s_prior}--\eqref{eq:sigma_prior} as they suppress zero and infinity. The choice of parameters for the hyperpriors \eqref{eq:sum_to_zero_normal}--\eqref{eq:sigma_prior} are informed by the range of probable values for each hyperparameter. Specifically, the following events each have a 0.99 prior probability of occurring:
\begin{align*}
    -5.15 &< \mu_h < 5.15 & \text{for } h=1,\ldots,H,  \\
    0.32 &< s_h < 8.85 & \text{for } h=1,\ldots,H, \\
    0.54 &< \tau_h < 14.52 & \text{for } h=1,\ldots,H, \\
    0.11 &< \sigma < 2.90.
\end{align*}

\begin{table}
\centering
\begin{tabular}{lccc}
\toprule
& MCMC-Exact & MCMC-Approx & LC-Sampling \\ \midrule
Burn-in iterations & 1000 & 1000 & 2000 \\
Inference iterations & 1000 & 1000 & 20000 \\
Total wall time (min) & 76.3 & 7.4 & 54.2 \\
\bottomrule
\end{tabular}
\caption{\label{table:tseries_iter}\small Number of iterations per MCMC chain (5 chains), and the total computational time taken by MCMC-Exact, MCMC-Approx, and LC-Sampling for the time-series example.}
\end{table}

We perform inference using NUTS for MCMC-Exact and MCMC-Approx, and Algorithm~2 (main text) for LC-Sampling. We report the number of MCMC iterations used and the computational wall time for each method in Table~\ref{table:tseries_iter}. Since the hierarchical model introduces correlations between model parameters, we increase the number of MCMC iterations performed. LC-Sampling requires more iterations as there is strong dependence between $\mathbf{z}_i$ and $\mathbf{p}_i$. We set the value of $C_i$ from Algorithm 2 to $C_i=10n_i$. We thin the number of LC-Sampling inference samples to 1000 per chain for the ESS comparison to be fair.

\begin{figure}[t]
\centering
\includegraphics[width=0.75\textwidth]{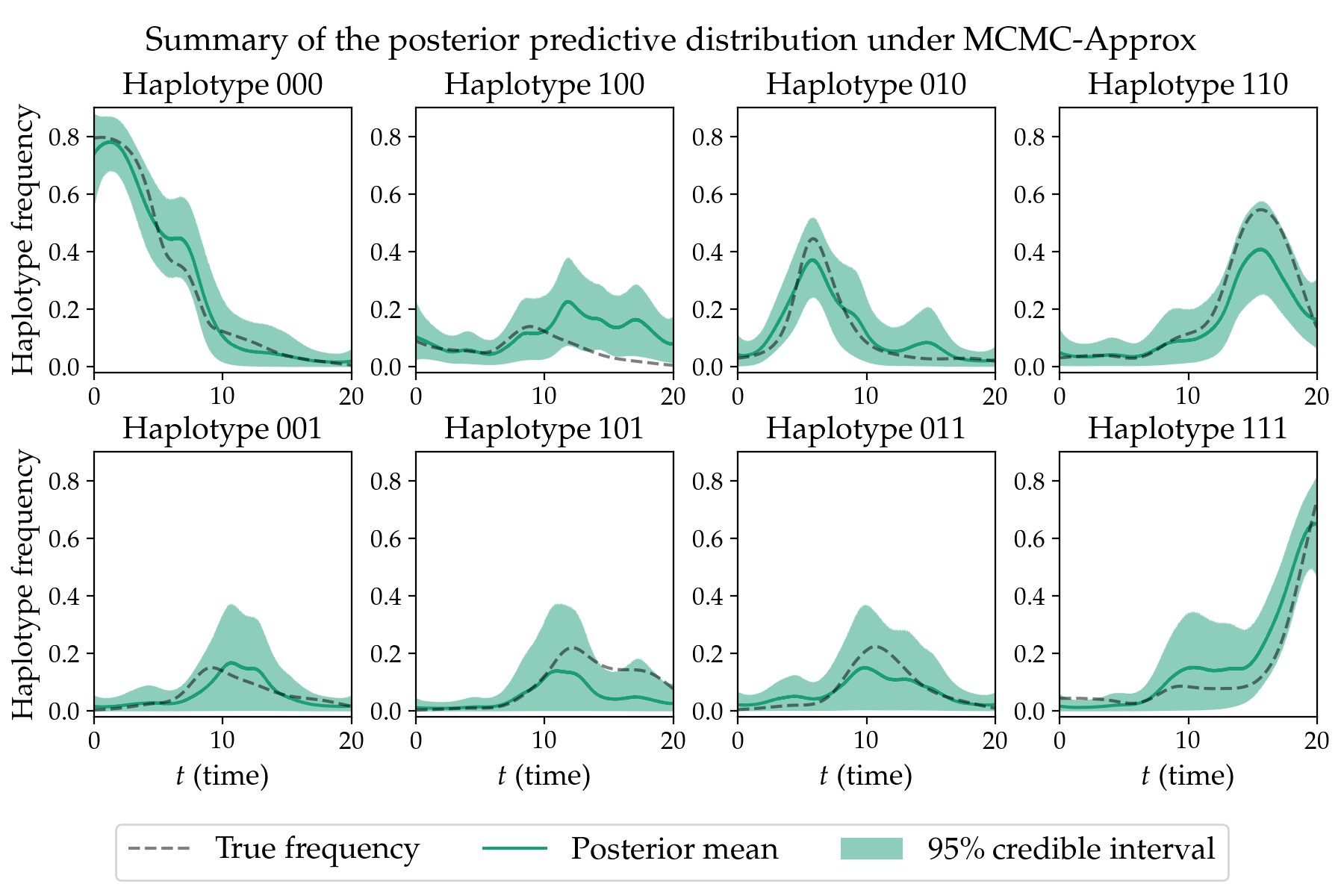}
\caption{\small Posterior predictive summary of haplotype frequencies under MCMC-Approx. The dashed and solid curves correspond to the true frequencies used for data simulation and the posterior mean respectively. Bands show 95\% credible intervals.} \label{fig:tseries_mn_trends}
\end{figure}

\begin{figure}[t]
\centering
\includegraphics[width=0.75\textwidth]{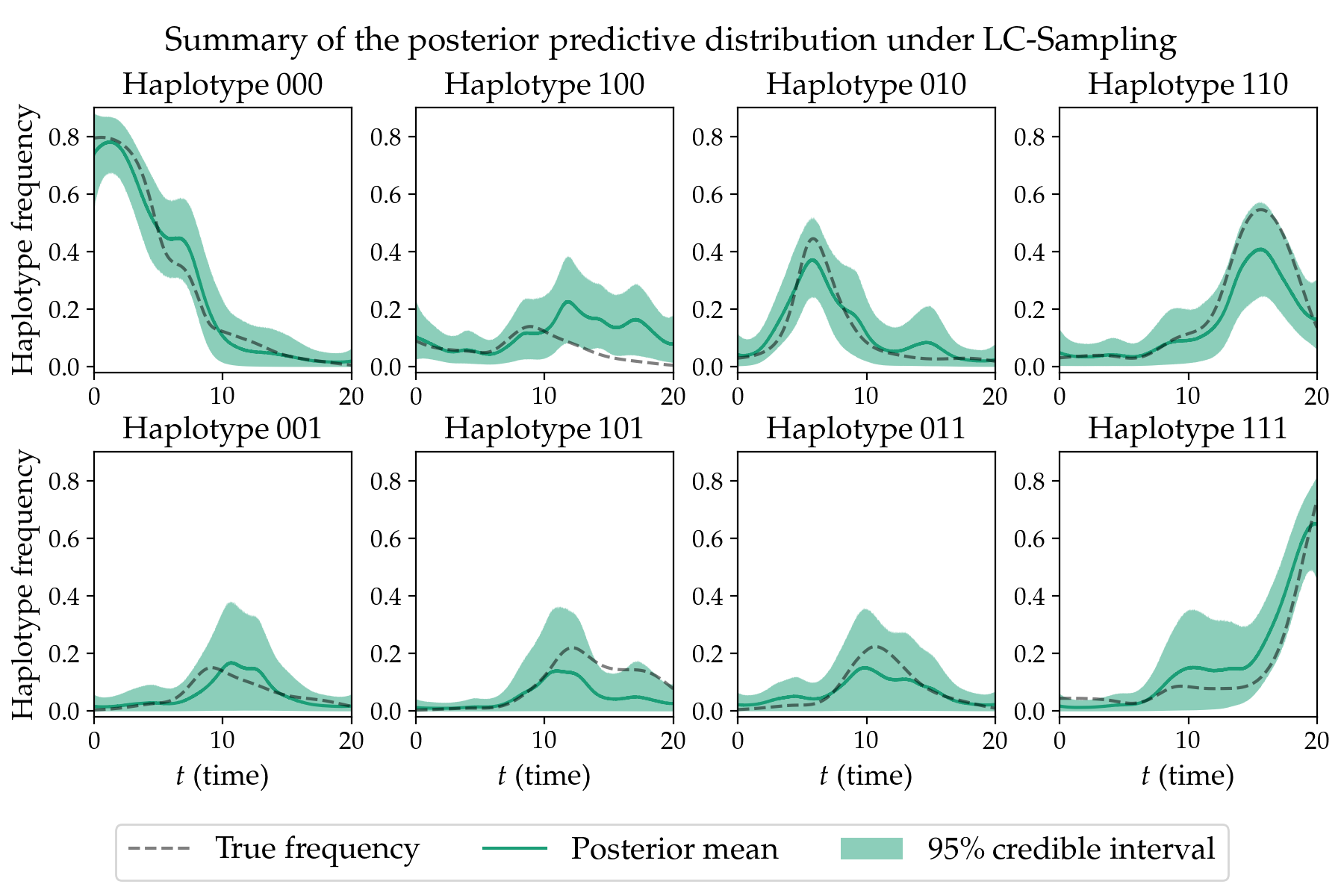}
\caption{\small Posterior predictive summary of haplotype frequencies under LC-Sampling. The dashed and solid curves correspond to the true frequencies used for data simulation and the posterior mean respectively. Bands show 95\% credible intervals.} \label{fig:tseries_gibbs_trends}
\end{figure}

\begin{figure}[t]
\centering
\includegraphics[width=0.56\textwidth]{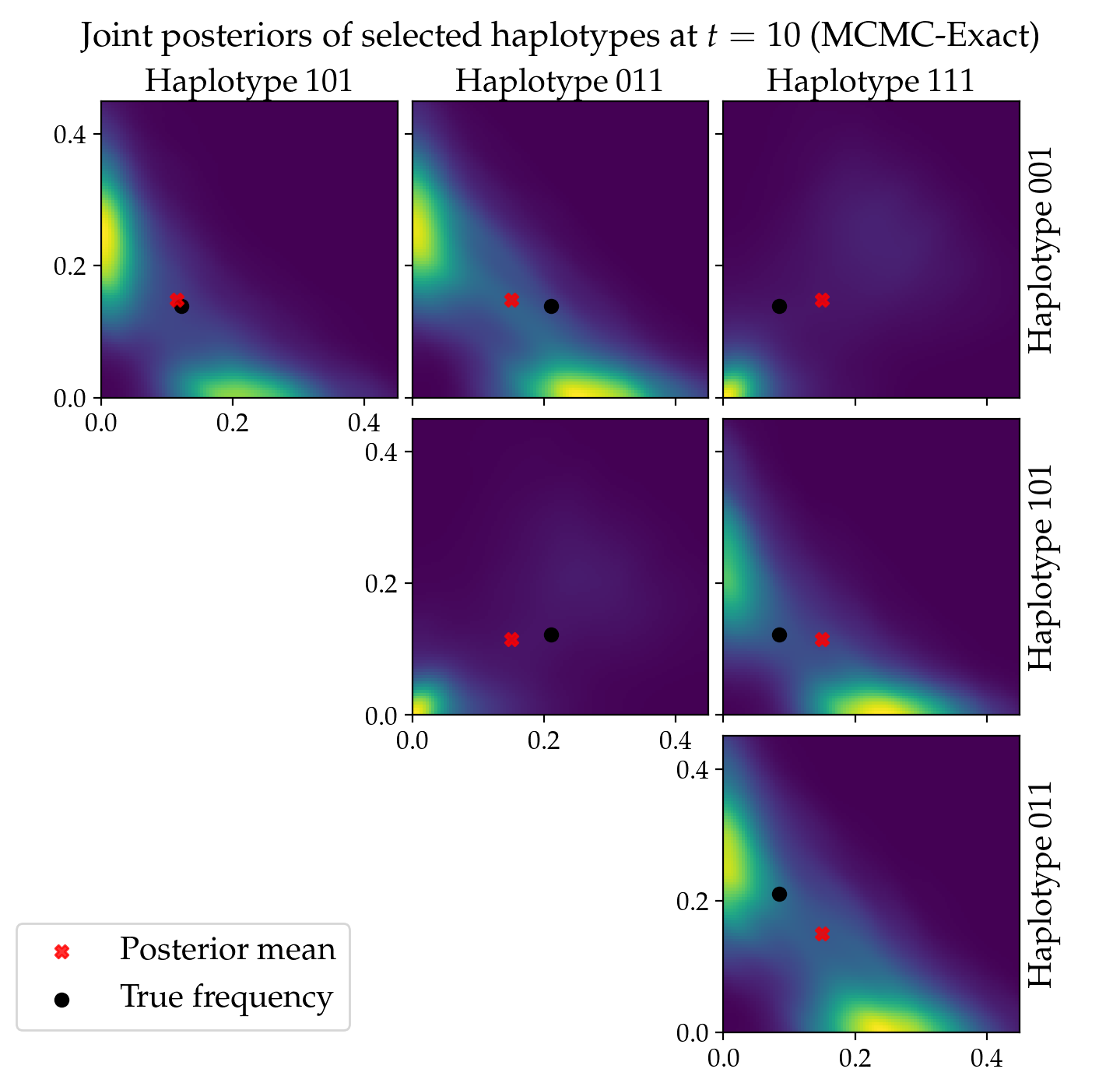}
\caption{\small Joint posterior distributions under MCMC-Exact of selected haplotype frequencies from the time-series example that show multimodality. The red cross and the black dot correspond to the posterior mean and the true frequencies respectively.}\label{fig:tseries_exact_joint}
\end{figure}

\begin{figure}[t]
\centering
\includegraphics[width=0.56\textwidth]
{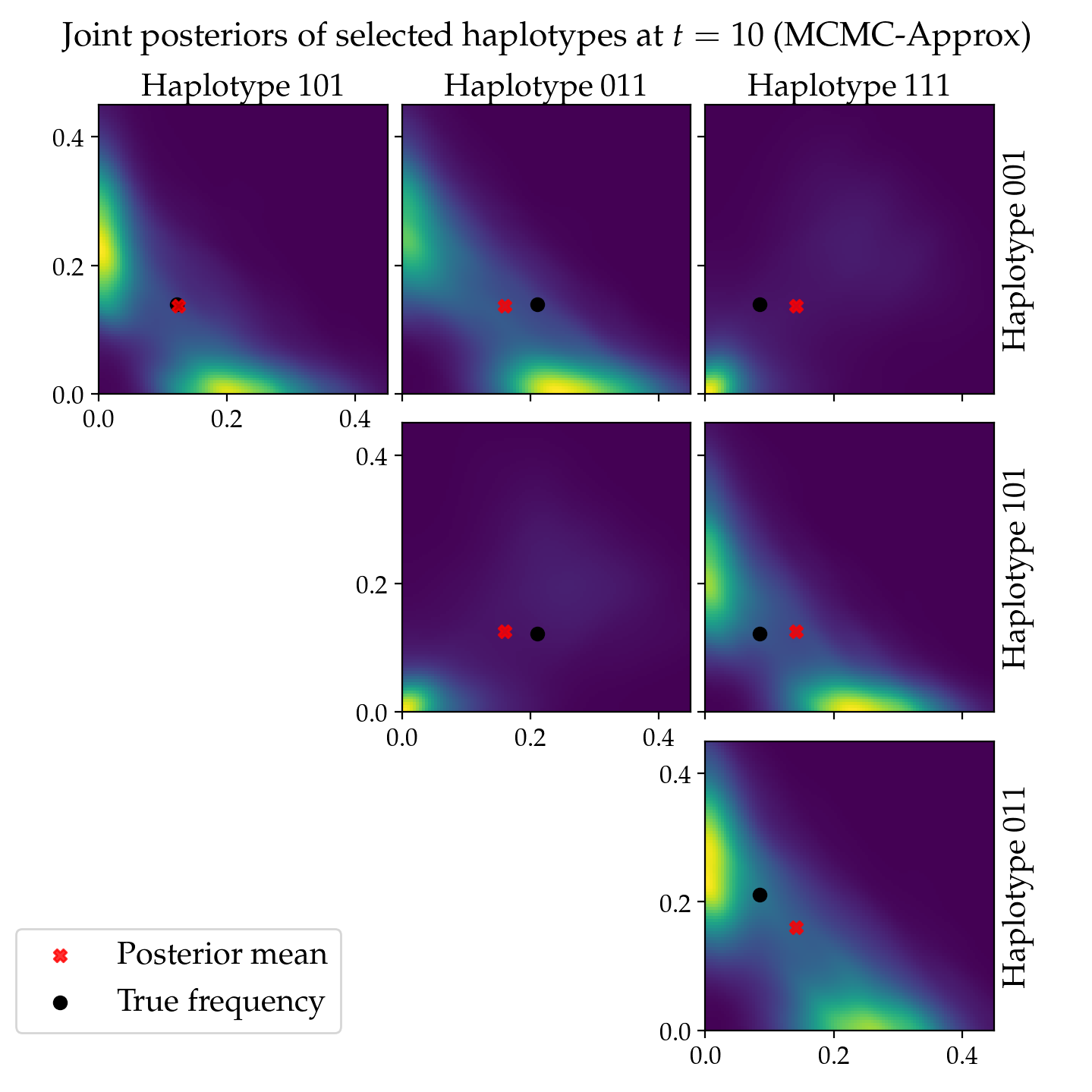}
\caption{\small Joint posterior distributions under MCMC-Approx of selected haplotype frequencies from the time-series example that show multimodality. The red cross and the black dot correspond to the posterior mean and the true frequencies respectively.}\label{fig:tseries_mn_joint}
\end{figure}

\begin{figure}[b]
\centering
\includegraphics[width=0.56\textwidth]
{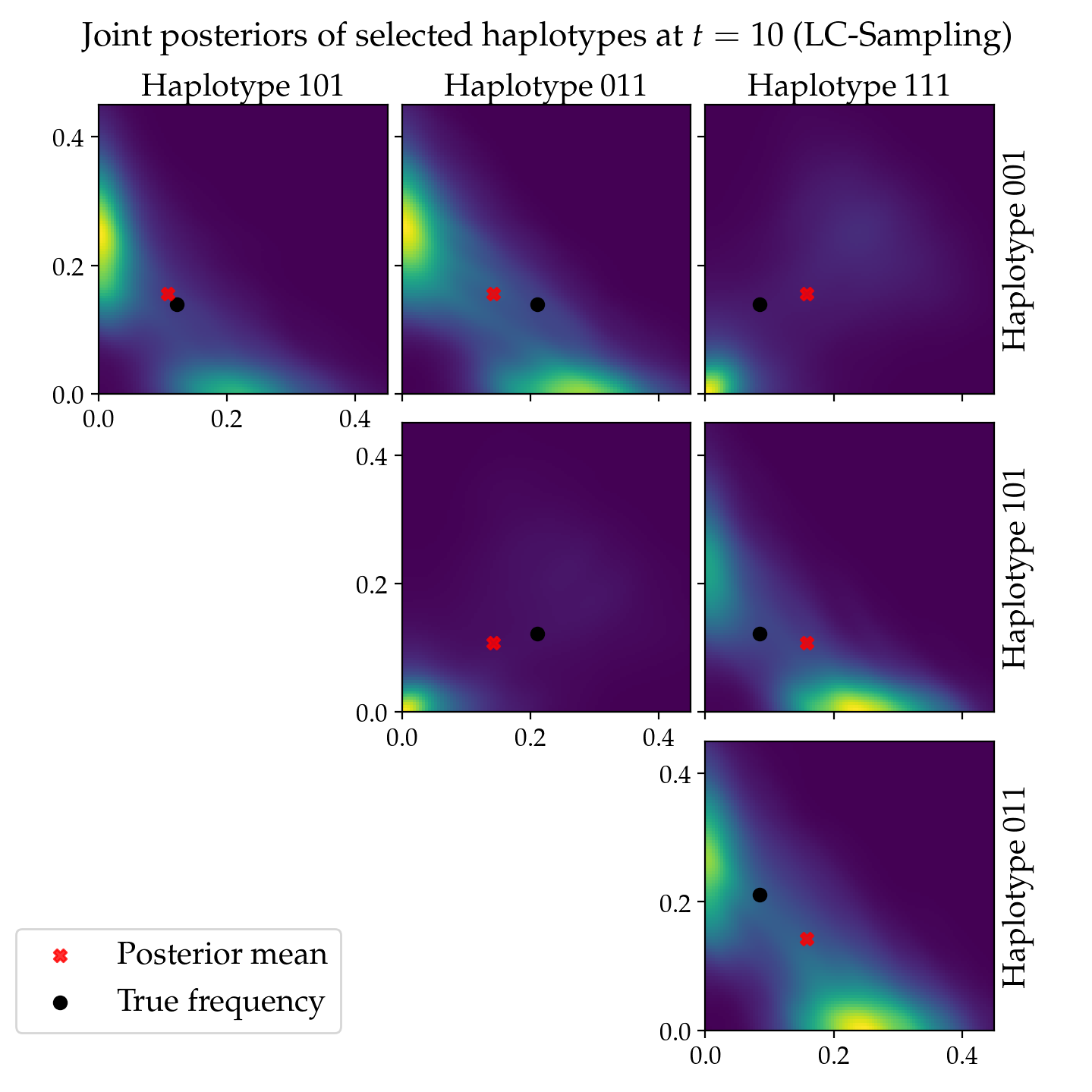}
\caption{\small Joint posterior distributions under LC-Sampling of selected haplotype frequencies from the time-series example that show multimodality. The red cross and the black dot correspond to the posterior mean and the true frequencies respectively.}\label{fig:tseries_gibbs_joint}
\end{figure}

To sample from the posterior predictive distribution of the haplotype frequency at any time $t$, we first sample the conditional normal distributions $f_h(t) \mid f_h(\textbf{t}),\mu_h,s_h,\tau_h,\sigma$ for each posterior sample of $\{f_h(\textbf{t}),\mu_h,s_h,\tau_h,\sigma\}$ over $h=1,\ldots,H$, then apply the softmax transformation to obtain
\begin{align}
p^\text{pred}_{h}(t) &= \frac{\exp(f_h(t))}{\exp(f_1(t))+\cdots+\exp(f_H(t))} & \text{for } \, h=1,\ldots,H.
\end{align}
The summaries of the univariate posterior predictive distributions for MCMC-Exact, MCMC-Approx, and LC-Sampling are shown in Figure~8 (main text), Figure~\ref{fig:tseries_mn_trends}, and Figure~\ref{fig:tseries_gibbs_trends} respectively. For the haplotypes 001, 101, 011, 111, there is multimodality in the posterior. As an example, we show the joint posterior distributions for these haplotypes at $t=10$ in Figures~\ref{fig:tseries_exact_joint}--\ref{fig:tseries_gibbs_joint}. For the joint distributions of haplotypes~101/011 and haplotypes~001/111, we observe a sharp mode near the origin (sparse frequencies), and a second mode with lower density and wider spread where the frequencies are away from zero. However, these two modes have comparable posterior mass as the posterior mean is located between the two modes. The other four joint distributions are characterised by a diagonal ridge. This suggests that we are able to infer the frequency of partial haplotypes where one of the first two markers does not have a specified allele (e.g. the partial haplotype ?01, which is 001 and 101 combined). However, there is a non-identifiability issue as there is insufficient signal in the data to infer the frequencies of the full haplotypes.

   
\printbibliography[heading=bibintoc]

@article{ito_estimation_2003,
	title = {Estimation of Haplotype Frequencies, Linkage-Disequilibrium Measures, and Combination of Haplotype Copies in Each Pool by Use of Pooled DNA Data},
	volume = {72},
	issn = {00029297},
	url = {https://linkinghub.elsevier.com/retrieve/pii/S0002929707605473},
	doi = {10.1086/346116},
	language = {en},
	number = {2},
	urldate = {2022-07-18},
	journal = {The American Journal of Human Genetics},
	author = {Ito, Toshikazu and Chiku, Suenori and Inoue, Eisuke and Tomita, Makoto and Morisaki, Takayuki and Morisaki, Hiroko and Kamatani, Naoyuki},
	month = feb,
	year = {2003},
	pages = {384--398},
}

@article{pirinen_estimating_2009,
	title = {Estimating population haplotype frequencies from pooled {SNP} data using incomplete database information},
	volume = {25},
	issn = {1367-4803, 1460-2059},
	url = {https://academic.oup.com/bioinformatics/article-lookup/doi/10.1093/bioinformatics/btp584},
	doi = {10.1093/bioinformatics/btp584},
	language = {en},
	number = {24},
	urldate = {2022-07-18},
	journal = {Bioinformatics},
	author = {Pirinen, M.},
	month = dec,
	year = {2009},
	pages = {3296--3302},
}

@article{pirinen_estimating_2008,
	title = {Estimating population haplotype frequencies from pooled {DNA} samples using {PHASE} algorithm},
	volume = {90},
	issn = {0016-6723, 1469-5073},
	url = {https://www.cambridge.org/core/product/identifier/S0016672308009877/type/journal_article},
	doi = {10.1017/S0016672308009877},
	language = {en},
	number = {6},
	urldate = {2022-07-18},
	journal = {Genetics Research},
	author = {Pirinen, Matti and Kulathinal, Sangita and Gasbarra, Dario and Sillanpää, Mikko J.},
	month = dec,
	year = {2008},
	pages = {509--524},
}

@article{iliadis_fast_2012,
	title = {Fast and accurate haplotype frequency estimation for large haplotype vectors from pooled {DNA} data},
	volume = {13},
	issn = {1471-2156},
	url = {https://bmcgenet.biomedcentral.com/articles/10.1186/1471-2156-13-94},
	doi = {10.1186/1471-2156-13-94},
	language = {en},
	number = {1},
	urldate = {2022-07-18},
	journal = {BMC Genetics},
	author = {Iliadis, Alexandros and Anastassiou, Dimitris and Wang, Xiaodong},
	month = dec,
	year = {2012},
	pages = {94},
}

@article{jajamovich_maximum-parsimony_2013,
	title = {Maximum-parsimony haplotype frequencies inference based on a joint constrained sparse representation of pooled {DNA}},
	volume = {14},
	issn = {1471-2105},
	url = {https://bmcbioinformatics.biomedcentral.com/articles/10.1186/1471-2105-14-270},
	doi = {10.1186/1471-2105-14-270},
	language = {en},
	number = {1},
	urldate = {2022-07-18},
	journal = {BMC Bioinformatics},
	author = {Jajamovich, Guido H and Iliadis, Alexandros and Anastassiou, Dimitris and Wang, Xiaodong},
	month = dec,
	year = {2013},
	pages = {270},
}

@article{patil_blocks_2001,
	title = {Blocks of limited haplotype diversity revealed by high-resolution scanning of human chromosome 21},
	volume = {294},
	issn = {0036-8075},
	doi = {10.1126/science.1065573},
	language = {eng},
	number = {5547},
	journal = {Science (New York, N.Y.)},
	author = {Patil, N. and Berno, A. J. and Hinds, D. A. and Barrett, W. A. and Doshi, J. M. and Hacker, C. R. and others},
	month = nov,
	year = {2001},
	pmid = {11721056},
	pages = {1719--1723},
}

@article{zhou_cshap_2019,
	title = {CSHAP: efficient haplotype frequency estimation based on sparse representation},
	volume = {35},
	issn = {1367-4803},
	shorttitle = {{CSHAP}},
	url = {https://doi.org/10.1093/bioinformatics/bty1040},
	doi = {10.1093/bioinformatics/bty1040},
	number = {16},
	urldate = {2023-06-19},
	journal = {Bioinformatics},
	author = {Zhou, Yinsheng and Zhang, Han and Yang, Yaning},
	month = aug,
	year = {2019},
	pages = {2827--2833},
}

@article{kirkpatrick_haplopool_2007,
	title = {{HAPLOPOOL}: improving haplotype frequency estimation through {DNA} pools and phylogenetic modeling},
	volume = {23},
	issn = {1367-4811},
	shorttitle = {{HAPLOPOOL}},
	doi = {10.1093/bioinformatics/btm435},
	language = {eng},
	number = {22},
	journal = {Bioinformatics (Oxford, England)},
	author = {Kirkpatrick, Bonnie and Armendariz, Carlos Santos and Karp, Richard M. and Halperin, Eran},
	month = nov,
	year = {2007},
	pmid = {17895275},
	pages = {3048--3055},
}

@article{link_uncovering_2010,
	title = {Uncovering a Latent Multinomial: Analysis of Mark-Recapture Data with Misidentification},
	volume = {66},
	issn = {0006341X},
	shorttitle = {Uncovering a Latent Multinomial},
	url = {https://onlinelibrary.wiley.com/doi/10.1111/j.1541-0420.2009.01244.x},
	doi = {10.1111/j.1541-0420.2009.01244.x},
	language = {en},
	number = {1},
	urldate = {2022-03-29},
	journal = {Biometrics},
	author = {Link, William A. and Yoshizaki, Jun and Bailey, Larissa L. and Pollock, Kenneth H.},
	month = mar,
	year = {2010},
	pages = {178--185},
}

@article{zhang_poool_2008,
	title = {{PoooL}: an efficient method for estimating haplotype frequencies from large {DNA} pools},
	volume = {24},
	issn = {1460-2059, 1367-4803},
	shorttitle = {{PoooL}},
	url = {https://academic.oup.com/bioinformatics/article-lookup/doi/10.1093/bioinformatics/btn324},
	doi = {10.1093/bioinformatics/btn324},
	language = {en},
	number = {17},
	urldate = {2022-07-18},
	journal = {Bioinformatics},
	author = {Zhang, Han and Yang, Hsin-Chou and Yang, Yaning},
	month = sep,
	year = {2008},
	pages = {1942--1948},
}

@Misc{4ti2,
  author = {{4ti2 team}},
  title =  {4ti2---A software package for algebraic, geometric and
            combinatorial problems on linear spaces},
  url =    {https://4ti2.github.io}
}

@article{Salvatier2016,
  doi = {10.7717/peerj-cs.55},
  url = {https://doi.org/10.7717/peerj-cs.55},
  year  = {2016},
  publisher = {{PeerJ}},
  volume = {2},
  pages = {e55},
  author = {John Salvatier and Thomas V. Wiecki and Christopher Fonnesbeck},
  title = {Probabilistic programming in Python using {PyMC}3},
  journal = {{PeerJ} Computer Science}
}

@article{kuk_computationally_2009,
	title = {Computationally feasible estimation of haplotype frequencies from pooled {DNA} with and without {Hardy}-{Weinberg} equilibrium},
	volume = {25},
	issn = {1367-4803, 1460-2059},
	url = {https://academic.oup.com/bioinformatics/article-lookup/doi/10.1093/bioinformatics/btn623},
	doi = {10.1093/bioinformatics/btn623},
	language = {en},
	number = {3},
	urldate = {2022-07-18},
	journal = {Bioinformatics},
	author = {Kuk, A. Y. C. and Zhang, H. and Yang, Y.},
	month = feb,
	year = {2009},
	pages = {379--386},
}

@article{zhang_fast_2019,
	title = {Fast likelihood‐based inference for latent count models using the saddlepoint approximation},
	volume = {75},
	issn = {0006-341X, 1541-0420},
	url = {https://onlinelibrary.wiley.com/doi/abs/10.1111/biom.13030},
	doi = {10.1111/biom.13030},
	language = {en},
	number = {3},
	urldate = {2021-04-28},
	journal = {Biometrics},
	author = {Zhang, W. and Bravington, M. V. and Fewster, R. M.},
	month = sep,
	year = {2019},
	pages = {723--733},
}

@article{hoffman_no-u-turn_2011,
  author  = {Matthew D. Hoffman and Andrew Gelman},
  title   = {The No-U-Turn Sampler: Adaptively Setting Path Lengths in {Hamiltonian} {Monte} {Carlo}},
  journal = {Journal of Machine Learning Research},
  year    = {2014},
  volume  = {15},
  number  = {47},
  pages   = {1593-1623},
  url     = {http://jmlr.org/papers/v15/hoffman14a.html}
}

@article{schofield_connecting_2015,
	title = {Connecting the latent multinomial: Connecting the Latent Multinomial},
	volume = {71},
	issn = {0006341X},
	shorttitle = {Connecting the latent multinomial},
	url = {https://onlinelibrary.wiley.com/doi/10.1111/biom.12333},
	doi = {10.1111/biom.12333},
	language = {en},
	number = {4},
	urldate = {2022-04-01},
	journal = {Biometrics},
	author = {Schofield, Matthew R. and Bonner, Simon J.},
	month = dec,
	year = {2015},
	pages = {1070--1080},
}

@article{niu_bayesian_2002,
	title = {Bayesian Haplotype Inference for Multiple Linked Single-Nucleotide Polymorphisms},
	volume = {70},
	issn = {00029297},
	url = {https://linkinghub.elsevier.com/retrieve/pii/S0002929707612907},
	doi = {10.1086/338446},
	language = {en},
	number = {1},
	urldate = {2022-07-15},
	journal = {The American Journal of Human Genetics},
	author = {Niu, Tianhua and Qin, Zhaohui S. and Xu, Xiping and Liu, Jun S.},
	month = jan,
	year = {2002},
	pages = {157--169},
}

@article{stephens_comparison_2003,
	title = {A Comparison of Bayesian Methods for Haplotype Reconstruction from Population Genotype Data},
	volume = {73},
	issn = {00029297},
	url = {https://linkinghub.elsevier.com/retrieve/pii/S0002929707619788},
	doi = {10.1086/379378},
	language = {en},
	number = {5},
	urldate = {2022-07-15},
	journal = {The American Journal of Human Genetics},
	author = {Stephens, Matthew and Donnelly, Peter},
	month = nov,
	year = {2003},
	pages = {1162--1169},
}

@article{diaconis_algebraic_1998,
	title = {Algebraic algorithms for sampling from conditional distributions},
	volume = {26},
	issn = {0090-5364},
	url = {https://projecteuclid.org/journals/annals-of-statistics/volume-26/issue-1/Algebraic-algorithms-for-sampling-from-conditional-distributions/10.1214/aos/1030563990.full},
	doi = {10.1214/aos/1030563990},
	number = {1},
	urldate = {2022-04-03},
	journal = {The Annals of Statistics},
	author = {Diaconis, Persi and Sturmfels, Bernd},
	month = feb,
	year = {1998},
}

@article{gasbarra_estimating_2011,
	title = {Estimating Haplotype Frequencies by Combining Data from Large {DNA} Pools with Database Information},
	volume = {8},
	issn = {1545-5963},
	url = {http://ieeexplore.ieee.org/document/5291689/},
	doi = {10.1109/TCBB.2009.71},
	number = {1},
	urldate = {2022-07-26},
	journal = {IEEE/ACM Transactions on Computational Biology and Bioinformatics},
	author = {Gasbarra, D and Kulathinal, S and Pirinen, M and Sillanpaa, M J},
	month = jan,
	year = {2011},
	pages = {36--44},
}

@article{the_1000_genomes_project_consortium_global_2015,
	title = {A global reference for human genetic variation},
	volume = {526},
	issn = {0028-0836, 1476-4687},
	url = {http://www.nature.com/articles/nature15393},
	doi = {10.1038/nature15393},
	language = {en},
	number = {7571},
	urldate = {2023-02-20},
	journal = {Nature},
	author = {{The 1000 Genomes Project Consortium} and Auton, Adam and Abecasis, Gonçalo R. and Altshuler, David M. and Durbin, Richard M. and Abecasis, Gonçalo R. and others},
	month = oct,
	year = {2015},
	pages = {68--74},
}

@book{hartl_popgen,
	edition = {4},
	title = {A Primer of Population Genetics and Genomics},
	isbn = {978-0-19-886229-1},
	url = {https://oxford.universitypressscholarship.com/view/10.1093/oso/9780198862291.001.0001/oso-9780198862291},
	language = {en},
	urldate = {2022-04-23},
	publisher = {Oxford University Press},
	author = {Hartl, Daniel L.},
	year = {2020},
	doi = {10.1093/oso/9780198862291.001.0001},
}

@article{liu_collapsed_1994,
	title = {The Collapsed {Gibbs} Sampler in {Bayesian} Computations with Applications to a Gene Regulation Problem},
	volume = {89},
	issn = {0162-1459},
	url = {https://doi.org/10.1080/01621459.1994.10476829},
	doi = {10.1080/01621459.1994.10476829},
	number = {427},
	urldate = {2023-06-06},
	journal = {Journal of the American Statistical Association},
	author = {Liu, Jun S.},
	month = sep,
	year = {1994},
	pages = {958--966},
}

@article{tam_benefits_2019,
	title = {Benefits and limitations of genome-wide association studies},
	volume = {20},
	copyright = {2019 Springer Nature Limited},
	issn = {1471-0064},
	url = {https://www.nature.com/articles/s41576-019-0127-1},
	doi = {10.1038/s41576-019-0127-1},
	language = {en},
	number = {8},
	urldate = {2023-08-09},
	journal = {Nature Reviews Genetics},
	author = {Tam, Vivian and Patel, Nikunj and Turcotte, Michelle and Bossé, Yohan and Paré, Guillaume and Meyre, David},
	month = aug,
	year = {2019},
	note = {Number: 8
Publisher: Nature Publishing Group},
	pages = {467--484},
}

@incollection{wright_genetic_2005,
	title = {Genetic Variation: Polymorphisms and Mutations},
	copyright = {Copyright © 2005 John Wiley \& Sons, Ltd. All rights reserved.},
	isbn = {978-0-470-01590-2},
	shorttitle = {Genetic {Variation}},
	url = {https://onlinelibrary.wiley.com/doi/abs/10.1038/npg.els.0005005},
	language = {en},
	urldate = {2023-08-06},
	booktitle = {Encyclopedia of {Life} {Sciences}},
	publisher = {John Wiley \& Sons, Ltd},
	author = {Wright, Alan F},
	year = {2005},
	doi = {10.1038/npg.els.0005005},
}

@inproceedings{hills_parameterization_1992,
	address = {Oxford},
	title = {Parameterization issues in {Bayesian} inference (with discussion)},
	booktitle = {Bayesian {Statistics} 4},
	publisher = {Oxford University Press},
	author = {Hills, S. E. and Smith, A. F. M.},
	editor = {Bernardo, J. M. and Berger, J. O. and Dawid, A. P. and Smith, A. F. M.},
	year = {1992},
	pages = {227--246},
}

@article{bonner_dmb,
	title = {Extending the latent multinomial model with complex error processes and dynamic {Markov} bases},
	volume = {10},
	issn = {1932-6157},
	url = {https://projecteuclid.org/journals/annals-of-applied-statistics/volume-10/issue-1/Extending-the-latent-multinomial-model-with-complex-error-processes-and/10.1214/15-AOAS889.full},
	doi = {10.1214/15-AOAS889},
	number = {1},
	urldate = {2022-05-15},
	journal = {The Annals of Applied Statistics},
	author = {Bonner, Simon J. and Schofield, Matthew R. and Noren, Patrik and Price, Steven J.},
	month = mar,
	year = {2016}
}

@article{hazelton_dmb,
	title = {Geometrically aware dynamic {Markov} bases for statistical linear inverse problems},
	volume = {108},
	issn = {0006-3444, 1464-3510},
	url = {https://academic.oup.com/biomet/article/108/3/609/5918020},
	doi = {10.1093/biomet/asaa083},
	language = {en},
	number = {3},
	urldate = {2022-05-15},
	journal = {Biometrika},
	author = {Hazelton, M L and Mcveagh, M R and van Brunt, B},
	month = aug,
	year = {2021},
	pages = {609--626},
}

@article{sibley_sp,
	title = {Pyrimethamine–sulfa\-doxine resistance in {Plasmodium} falciparum: what next?},
	volume = {17},
	issn = {14714922},
	shorttitle = {Pyrimethamine–sulfadoxine resistance in {Plasmodium} falciparum},
	url = {https://linkinghub.elsevier.com/retrieve/pii/S1471492201020852},
	doi = {10.1016/S1471-4922(01)02085-2},
	language = {en},
	number = {12},
	urldate = {2022-04-23},
	journal = {Trends in Parasitology},
	author = {Sibley, Carol Hopkins and Hyde, John E and Sims, Paul F.G and Plowe, Christopher V and Kublin, James G and others},
	year = {2001},
	pages = {582--588},
}

@article{ebel_angola,
	title = {Historical trends and new surveillance of {Plasmodium} falciparum drug resistance markers in {Angola}},
	volume = {20},
	issn = {1475-2875},
	url = {https://malariajournal.biomedcentral.com/articles/10.1186/s12936-021-03713-2},
	doi = {10.1186/s12936-021-03713-2},
	language = {en},
	number = {1},
	urldate = {2021-04-13},
	journal = {Malaria Journal},
	author = {Ebel, Emily R. and Reis, Fátima and Petrov, Dmitri A. and Beleza, Sandra},
	month = dec,
	year = {2021},
	pages = {175}
}
\end{refsection}
\end{appendices}

\end{document}